\documentclass[epj]{svjour}

\usepackage{amsmath,amssymb}
\usepackage{graphicx,url}
\begin{document}

\title{Modern temporal network theory: A colloquium}

\author{Petter Holme\inst{1}}

\institute{Department of Energy Science, Sungkyunkwan University, Suwon 440-746, Korea}
\mail{holme@skku.edu}
\date{Received: date / Revised version: date}

\abstract{
The power of any kind of network approach lies in the ability to simplify a complex system so that one can better understand its function as a whole. Sometimes it is beneficial, however, to include more information than in a simple graph of only nodes and links. Adding information about times of interactions can make predictions and mechanistic understanding more accurate. The drawback, however, is that there are not so many methods available, partly because temporal networks is a relatively young field, partly because it more difficult to develop such methods compared to for static networks. In this colloquium, we review the methods to analyze and model temporal networks and processes taking place on them, focusing mainly on the last three years. This includes the spreading of infectious disease, opinions, rumors, in social networks; information packets in computer networks; various types of signaling in biology, and more. We also discuss future directions.
}

\PACS{
      {89.75.Hc}{Networks and genealogical trees} \and
      {02.10.Ox}{Combinatorics; Graph theory}   \and
      {89.70.-a}{Information and communication theory} }

\maketitle

\section{Introduction}

To understand how large connected systems works, one needs to zoom out and view them from a
distance. In other words, one needs a principled, consistent way of discarding irrelevant information. A common way of doing this is to represent the system as a network, where nodes are connected if they interact. For many systems one  have more information that just about who interacts. Including that information into a \emph{temporal network}, of course, goes against the idea of simplifying the system. Sometimes, however, it could be worth the effort in terms of increased accuracy of predictions, increased mechanistic understanding, etc. The drawback is that many of the methods and models developed for static networks could be inapplicable or could need non-trivial generalizations.

Introducing temporal networks like above, as an extension of static networks is natural. After all, they are based on a mathematical structure that links entities pairwise (only that they also encode the time of the interaction). However, there are surprisingly deep differences, both in the history of the fields, the methods used and the questions asked.
For static networks~\cite{newman_book,barabasi_book}, many of the central concepts were developed to study social networks---centrality measures, community detection, similarity measures, etc.\ all come from network sociology. As more disciplines came to embrace networks, these concepts were either reinvented or adopted from social network studies~\cite{wasserman_faust}. In a similar fashion, computer science pioneered temporal network theory. Still today, researchers rediscover the ideas Leslie Lamport and others used in the 1970's to build a theory of distributed computing~\cite{lamport}.

At a very fundamental level, the mathematics of temporal and static networks differ. We will refer to the basic unit of interaction in a temporal network as a \emph{contact}. It captures information about a pair of nodes interacting and the time of the interaction. A contact is the closest extension of a \emph{link} in static networks (but we will reserve link for a static relationship between two nodes---usually that they have one or more contacts). Being connected is a transitive mathematical relation, i.e.\ if $(i,i')$ and $(i',i'')$ are links then $i$ is  connected through a path. This is true also for directed static networks, but does not have to be true for contacts in a temporal network. As a corollary, there is no way of representing a temporal network as a simple graph without either losing information or changing the meaning of the nodes.  Because of this and other reasons, the analysis of temporal networks has become rather different than static networks. Some aspects (like visualization) are much harder and less fruitful than for temporal networks; other techniques (like randomized null-models) are richer and more powerful for temporal networks.

Temporal networks is a very interdisciplinary field. Computer scientists, physicists, mathematicians, engineers, social scientists, medical researchers and the occasional biologist all helped shape it to what it is. The problems addressed reflect this diversity. They range from classic computer science questions (like constrained optimization~\cite{fomin}, the traveling salesman problem~\cite{tsp} or constructing minimal spanning trees~\cite{huang_mst}) to physics-influenced papers about phase transitions~\cite{posfai_phasetransition}. They cover systems from cattle in farms~\cite{chen_animal} to citations between scientific papers~\cite{clough_trans}. This diversity has boosted the development and one may wonder how many other disciplines that reached maturity this fast. The flip-side is that many concepts have been rediscovered, or developed in parallel. No wonder perhaps, the terminology is in disarray. Even the topic of the field itself goes under many names---time-varying networks, dynamic networks and temporal graphs. Preparing this colloquium paper, we planned to make a glossary to translate between papers, but soon we came to realize that it is better not to lead the reader into false security. Just be aware that when you see a paper, for example, ``temporal distance'' it could mean at least three different things, and look up the definitions paper by paper. We will try to keep the terminology from Ref.~\cite{holme_saramaki_rev} as much as possible, including that our default type of temporal network is a stream of instantaneous contacts between two nodes in discrete time. The second confusing situation (that we, on the other hand, will try to sort out) is that there are subtly different ways of thinking of temporal networks---are they streams of contacts or static links that are occasionally active? These mental pictures could be mathematically equivalent but still shape the research profoundly. 

In this colloquium, we will give a general introduction to temporal networks with an overview of the methods, systems modeled and questions addressed. Specifically, we focus on the development after our review paper Ref.~\cite{holme_saramaki_rev}. We do not cover studies of network evolution (where the network is well-approximated as static for dynamic systems confined to it) or algorithmic papers that do not aim at understanding real systems. We will not discuss \emph{adaptive networks}~\cite{gross_adaptive,egui_ajs,wardil_hauert} much either. These are networked systems with a feedback between the topology and a dynamic system on the network. They differs from temporal networks by the time of events being of subordinate importance to topology. In practice temporal network studies have a stronger connection to empirical data and adaptive network papers are mostly purely modeling studies.

\section{Systems that can be modeled as temporal networks}

In this section, we will present the systems that people have modeled as temporal networks. There are probably many systems that will be added to this list in the near future. In general, any system with pairwise interactions and information about time could be modeled as a temporal network. Simply speaking, for such modeling to be practically useful, there should be some structure in both time and network topology that affects  dynamic systems on the network. This is usually equivalent to saying that the time scale of the dynamics of the network should not be too far from the time scale of the dynamic system.

\subsection{Human proximity networks} \label{sec:proxi}

One of the most well-studied type of temporal networks (but still far from fully understood). Is that of human proximity networks. These are networks that record when two people have been close to each other in time and space. Researchers have been very creative to measure this type of interactions. The highest resolution data we are aware of comes from radio-frequency identification (RFID)~\cite{thebook:barrat,f2f2,kibanov,cattutto_neo4j,panisson_fingerprinting,gallery,conference,sociopatterns_hospital0,sociopatterns_hospital1} or infrared~\cite{taro_prx} sensors that can measure when people are within a few meters apart and facing each other. Typically one filters out very brief contacts, so the remaining contacts capture people either in a face-to-face conversation or in a crowd or gathering of some kind. WiFi~\cite{salathe,toth,hornbeck} or Bluetooth~\cite{arek_vedran,vedran_arek_sune,issue:sune,pfitzner_betweenness,scholtes_causality} sensors give almost the same resolution (with a few meters lower positional accuracy but worse angular resolution). Of these references, Refs.~\cite{cattutto_neo4j,arek_vedran} are probably the most practically useful for those wanting to set up their own studies. Zhang et al.~\cite{li_a} reviews many aspects of  temporal networks of human proximity.

The drawback with the above type of high-resolution data sets is that such studies are time consuming and costly. To sample larger populations, one need to give up the precision. One example of such include the studies of Zhang, Li et al.~\cite{li0,li1} of people logged onto a campus WiFi networks (where a contact is recorded if two persons are connected to the same WiFI router at the same time). Similarly Yoneki, Hui and Crow\-croft~\cite{yoneki} use a network of people found to be close in space and time by stationary scanners for Bluetooth hardware addresses. In another large-scale low-resolution study Sun et al.~\cite{sun_axhausen} study commuters in Singapore. Here contacts were recorded for passengers on the same bus. Kim, Lee and Shroff~\cite{opportunistic} use a network of spatiotemporal proximity of taxis in Shanghai (probably not so interesting as far as human-to-human spreading or diffusion processes goes, but perhaps for other types of information spreading).

A special type of human proximity networks that have got much attention is patient-referral networks, i.e.\ how patients are transferred between wards of a hospital system. Ref.~\cite{liljeros_giesecke_holme} was the first such study, to our knowledge, studying 295,108 Swedish patients over two years. This paper discusses how to best reduce the temporal network  to a static network (more about this in Sec.~\ref{sec:window}). Walker et al.~\cite{walker_cdiff} study the spread of \textit{C. difficile} among 166,542 patients for 2.5 years. Donker et al.~\cite{donker_wallinga_grundmann} study a one-year data set of 1,676,704 patients in the Netherlands that they, for privacy reasons, reduce to a metapopulation model. There are also smaller-scale, higher-quality studies using sensors rather than healthcare records, as mentioned above~\cite{sociopatterns_hospital0,sociopatterns_hospital1,hornbeck}.

Yet another kind of human proximity network (perhaps so different that it needs a category of its own) are networks of sexual contacts. These have the same quality vs.\ quantity problem as mentioned before. Classic sexual network studies like Refs.~\cite{potterat,haraldsdottir} do not have the time of the contacts. The only large-scale temporal network of sexual contacts is the prostitution data of Rocha et al.~\cite{rocha_liljeros_holme} where the sexual contacts are self-reported by Brazilian sex buyers at a web community.

Several proximity networks are publicly available (analyzing these is probably the best way to get started in temporal network research). See the supplementary material in Refs.~\cite{salathe,rocha_comp_biol} and \url{http://www.sociopatterns.org/}.

\subsection{Animal proximity networks} \label{sec:aniprox}

Researchers have not only been interested in the temporal proximity networks of humans. There is a relatively large number of studies of temporal networks of animals. Mostly populations of livestock modeled either as a metapopulation networks (where one farm is one node and a transport between two farms is a directed contact), or as a temporal network of individuals where a contact represents being at the same farm at the same time. Livestock here could refer to either cattle~\cite{gates_woolhouse,genois_vacc,scholtes_causality,valdano_poletto,chen_animal} or swine~\cite{konschake_components}. Most such networks are inferred by the records of the farms, and regulating authorities, but Ref.~\cite{chen_animal} use RFID devices attached to the ears of calves. Not only domesticated, but also wild animals have been studied---zebras~\cite{lahiri_berger_wolf_2007,sulo} and capuchin monkeys~\cite{capuchin} by GPS traces, and ants~\cite{thebook:charbonneau} by visual observation. Psorakis et al.~\cite{psorakis} study networks of great tits from foraging records.

\subsection{Human communication}

Temporal networks of human communication maybe needs no further introduction as another \textit{Eur.\ Phys.\ J.\ B} colloquium covers these~\cite{jari_moro}. They are together with proximity networks the largest class of systems modeled as temporal networks. A typical kind of such data comes from call-data records of mobile phone operators~\cite{mantegna,bernhardsson,kivela_multiscale,karsai_slow_small,kovanen_motif,thebook:miritello,miritello_limited,miritello_dyn_strength,jiang_calling}. These register who called whom, or who sent text messages to whom, typically restricted to one operator in one country. Another type of communication networks are e-mails. These data sets have been sampled from the email accounts of a group of people. When using this data to create a static network the boundary condition is major problem---should one include e-mails outside of the group~\cite{ebel} or not~\cite{eckmann}? This is perhaps an even larger problem for temporal-network studies of large social media platforms such as Twitter~\cite{FerrazCosta,romero,sanli}. For some temporal network studies focusing on  individuals, this becomes less of a problem, since one have all actions recorded of the sampled people~\cite{barabasi_bursts}. The immediate boundary problem more or less disappears if one studies complete records of closed Internet communities or social networking services~\cite{holme_edling_liljeros,karimi_ramenzoni,villani_frigessi,jacobs_thefacebook,mathiesen,karsai_skype}, but even for these cases the dynamics is of course also shaped by events and communication outside of the system. 

\subsection{Collaboration networks}

A well-studied system in static networks that is naturally time resolved is collaboration networks. In these, a contact represents that people do something together, but not necessarily at the same place (like proximity networks). Scientific collaborations are a particularly well studied topic, especially for static networks. Pfitzner et al.~\cite{pfitzner_betweenness} Moinet et al.~\cite{moinet} is one exception in the temporal network literature, but note that early static network papers like~\cite{mejn_prer} also touch upon some temporal aspects (even though the main question is why the static network of accumulated contacts look like it does).

\subsection{Citation networks}

Another type of network that is well studied in the static network literature is citation networks. Time puts a strong constraint on the static structure of these networks in that they have to be acyclic~\cite{karrer_dag,wu_holme}. In other words, you have to be able to order the nodes in such a way that all the directed links between them point in one direction. They are also special in the sense that all out-links of a node (paper) happen simultaneously (when the paper is published). Rosvall et al.~\cite{rosvall_memory,rosvall_alluvial}  and Clough et al.~\cite{clough_trans} discuss temporal aspects of this type of data.

\subsection{Economic networks} \label{economy}

Economic networks are perhaps a bit understudied as temporal networks. Petri and Expert studies a more than one century long data set of trade between countries~\cite{petri_expert}. Kondor et al.~\cite{kondor_posfai} analyze Bitcoin transactions (with more standard network approach than we discuss in this paper, but the data itself is a temporal network). Tan et al.~\cite{creditcard} study a temporal network of credit card transactions. Redmond and Cunningham investigate an online site that administrates direct loans between the members. Xin et al.~\cite{xin_motif} study temporal network data sets of the ship chartering and build-to-order ship markets. Another paper by Xin et al.~\cite{issue:xin} analyze the Chinese venture capital market. Popovi\'c et al.~\cite{term_co} consider networks of countries inferred from financial news and compare them to correlation networks from  time series of the price of credit default swaps. In principle, many of the other network data sets studied in the (static) network literature~\cite{catanzaro,econnwk} could be understood by temporal-network methods. 

\subsection{Brain networks} \label{sec:brain}

In neuroscience, networks have become a useful tool to understand the organization function of the brain and how different types of conditions alter the coupling between different regions~\cite{park_friston,sporns,baronchelli_cognitive}. The most common type of network is constructed from temporal correlations of the oxygen levels as measured by fMRI scanning. Even though fMRI has a temporal resolution of the shortest time scale of neuronal activities, it has proven fruitful to study as a temporal network~\cite{bassett_brain,bassett_learning,mantzaris_brain}.

\subsection{Travel and transportation networks}

Networks of human transportation systems lends themselves well to a temporal-network modeling framework. We already mentioned Sun et al.'s study of the bus transportation in Singapore~\cite{sun_axhausen}. Scholtes et al.~\cite{scholtes_causality} and Rosvall et al.~\cite{rosvall_memory} study networks of airline connections. Scholtes et al.~\cite{scholtes_causality} also study a temporal network of subway travel in London. Kaluza et al.~\cite{kaluza} investigate the network of global ship transport. One could in principle let the nodes represent vehicles instead of people (although we are not aware of such studies). On a smaller scale, Borgnat et al.~\cite{borgnat_velov} study a shared-bike system in a French city. In the interface of transportation studies and ecology Banks et al.~\cite{banks_eco} study how transportation networks help the migration of species.

\subsection{Distributed computing}

Many of the concepts that we discuss in this paper were, as mentioned, first developed in the theory of distributed computing systems~\cite{lamport,kuhn}. After a slow start, the interest in this area has increased a lot. One reason is the development of cheap wireless devices, another is that there are many theoretical challenges (especially when the units are moving around in space or in and out of a network). The goals for such systems are usually (quoting Michail~\cite{michail_temporal_graphs}) ``to compute (i.e.\ agree on) something useful or construct a desired network or structure in such an adversarial setting.'' Another direction in this area is to determine the condition for desirable properties to hold for a distributed computing system under the dynamics (``churn'') of the devices. These studies are to our knowledge all theoretical. It would be interesting to see more empirical temporal network studies of distributed computing systems.

\subsection{Ecological networks} \label{sec:econwk}

In ecology, networks have mostly been used to capture interactions between species~\cite{pascualdunne,sole_bascompte}. Food webs, is a typical example of an antagonistic interaction between species, describing what species that eats what other species. There are also mutualistic interactions where both interacting species benefit from the interaction (plants and pollinators being a typical example~\cite{rasmussen}). Both these types of networks change with the season (and also from longer-term effects due to climate change etc.)\ and could therefore be studied with the methods described in this paper. Another example of ecological networks that could be, but to our knowledge has not yet been, studied are interlinked habitats~\cite{matisziw,hobbs}---the underlying structure for meta-community studies of the process of colonization and extinction in networks of habitat patches. In a long term, this network also changes in time. A third class of potentially interesting networks in (behavioral) ecology are networks of individual animals. This overlaps with the animal proximity networks discussed above, but the interaction could be more indirect~\cite{hasenjager}. An introduction of temporal networks for ecologists and evolution theorists can be found in Ref.~\cite{blonder_rev}.

\subsection{Biological networks}

In biology and the `omics there are also plenty of systems that could be modeled as temporal networks. Static network modeling has mainly been applied to gene networks, protein-interaction networks and metabolic networks. Gene networks can capture many different types of interactions---from regulatory networks (where a link means that one gene activates or inhibits another), via gene-fusion networks (of genes that can form hybrid genes), to more abstract relationships between genes also including information of their encoded proteins interact and their distance to each other on the DNA~\cite{string}. Researchers have studies some kind of temporal effects of all these three levels of networked `omes, but it is fair to say adding the dimension of time has been harder than for some of the examples above. One cannot yet record when a reaction happens in a metabolic network, at least on a large scale, or when two proteins bind to each other. However, with future improved technologies this could change.

We will list a few examples of network biology including time in the modeling. For none of these a contact is not as precisely recorded as e.g.\ human proximity data. Kharchenko et al.~\cite{kharchenko} discuss the genetic control of metabolism, but the temporal component comes from modeling, not measurement. Gyurk\'o et al.~\cite{gyurk} discuss the use of networks of individual proteins to understand the development of cancer. Taylor et al.~\cite{taylor} argue that temporal reorganizations of the protein interaction network (the network of proteins that \emph{do} interact, not the network of proteins that \emph{could} interact) could predict and explain the development of breast cancer. Luo et al.~\cite{luo_new} predict essential proteins by temporal networks. Rigbolt et al.~\cite{rigbolt} discuss temporal aspects of gene networks in cell differentiation. As a final example, West et al.~\cite{west_bianconi} investigate the use of a temporal network entropy~\cite{thebook:bianconi} to understand the evolution of cancer.

\subsection{Other systems}

The topics above are by no means all the possible temporal networks to be studied. We believe the readers of this paper are more imaginative that the author, so we will not try much harder. To mention two more systems, in one extreme Ronhovde et al.~\cite{ronhovde} study temporal networks of glassy states in complex materials, in the other extreme are narrative networks (telling a story about a complex, interdependent set of events)~\cite{bear_hist,Beek92reasoningabout}.

\begin{figure}
\begin{center}
\includegraphics[width=0.65\linewidth]{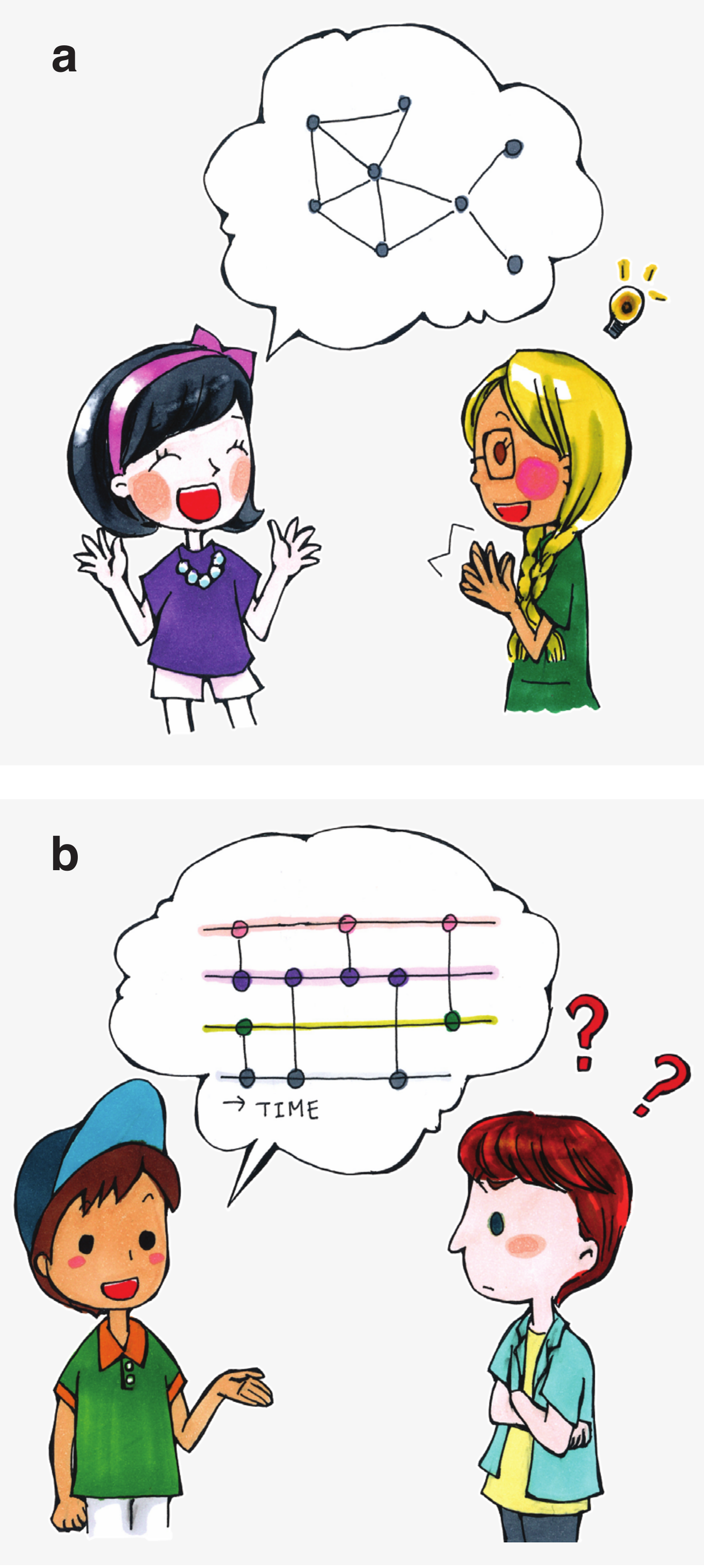}
\end{center}
\caption{Illustration of a main challenge for temporal networks. One of the great benefits of static networks (a) is that they are great to support discussions of how a system is connected, and how dynamics on top of it behave. Temporal networks do not have this graphical simplicity, and are thus much less intuitive (b). Illustration: Mi Jin Lee.}
\label{fig:graphical_mijin}
\end{figure}

\section{Representations of temporal networks}

One can incorporate information about the time of contacts between pairs of nodes in many ways. Which way to chose could depend of what one's data mean, how accurate it is, what type of process the temporal network supports, etc. Furthermore, it reflects the way the researcher conceptualizes her system, and is usually not well motivated in  papers, even though it hides many assumptions about the data. In this section, we will discuss some general ways of representing temporal networks that sometimes translate to a graphical representation (Fig.~\ref{fig:graphical_mijin})~\cite{batagelj} or a data structure, sometimes not.

\begin{figure*}
\begin{center}
\includegraphics[width=0.8\linewidth]{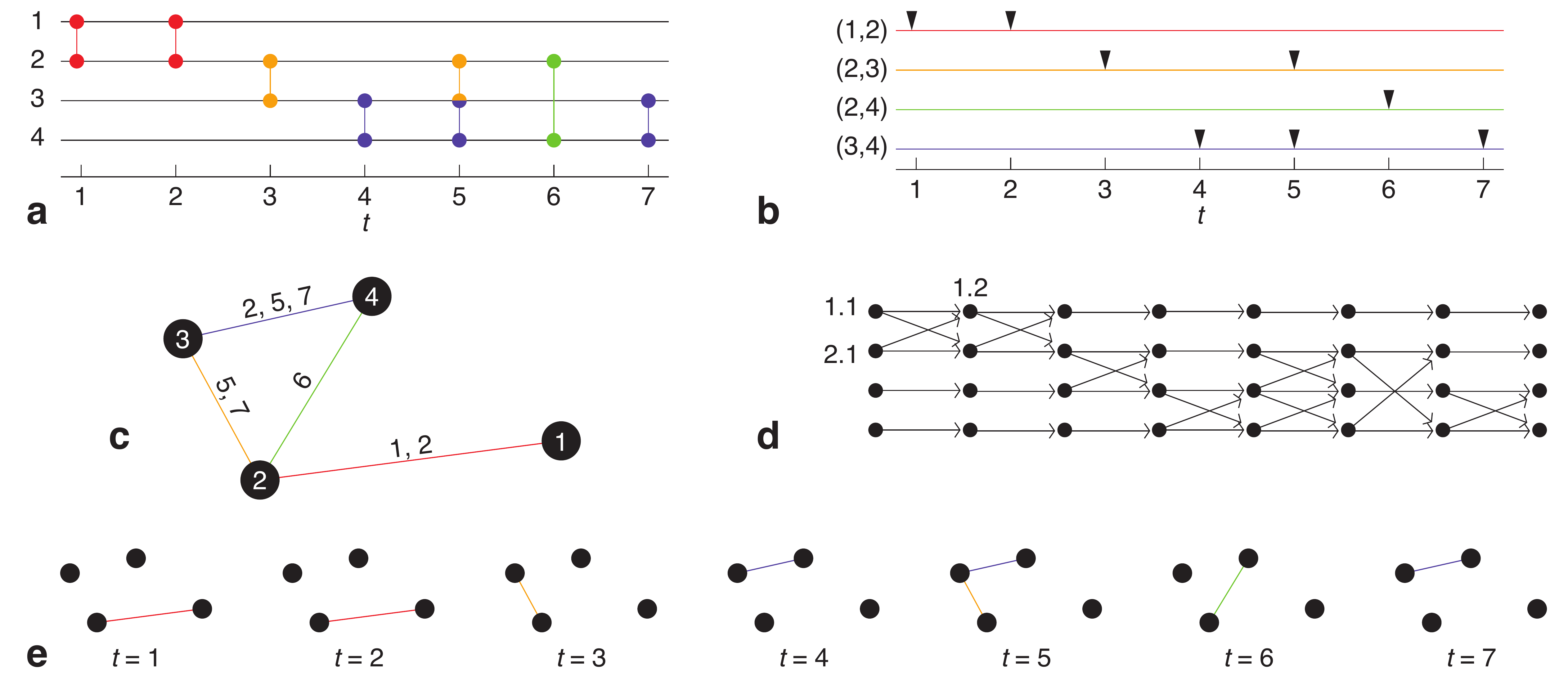}
\end{center}
\caption{The figure illustrates five representations of the same temporal network (of the contact sequence type). Panel (a) shows a node-centric time line, where a horizontal line represents a contact between two connected individual at the time given by the x-axis. Panel (b) shows a time line of the contacts focusing on links (pairs of individuals). Panel (c) shows a time-stamp decorated, aggregated graph. The numbers of the links denotes the contacts between the nodes. Panel (d) shows a time-node graph (where one assumes spreading cannot occur across more than one contact per time step). Three of the 32 time nodes are labeled. Panel (e) shows a graph sequence representation.}
\label{fig:lossless}
\end{figure*}

\subsection{Lossless representations}

We borrow the word ``lossless'' from audio compression to denote representations that can carry all information about a temporal network. They are all theoretically equivalent, but do shape the way to think about a temporal network.

\subsubsection{Contact sequences}

Almost all empirical temporal-network data sets that we have seen are of the form of lists of contacts, i.e.\ the two node involved and the time of the interaction---either just one time stamp (that also means that the time is discretized) or an interval (this case is sometimes called an \textit{interval graph}). This is a very easy  and practical format com\-pu\-ta\-ti\-o\-nally---just a list of three or four columns---nothing that would scare even an  Excel-only user. Just as good contact sequences are for computational purposes, just as bad are they to think about the function of the system or processes confined to the network. You would never see a temporal network researcher jotting down a contact sequence to discuss a new method. The reason is, of course, that contact sequences lack any kind of visual expression power.

\subsubsection{Graph sequences or multilayer networks} \label{sec:graph_seq}

Some authors present temporal networks as sequences of static graphs~\cite{michail_temporal_graphs,perra_activitydriven,nicozia_metrics}, or, equivalently, layers in a multilayer network~\cite{boccaletti_rev,kivela_rev,goh_multi}. Depending on the system studied, this could be a very powerful way to think about temporal network, simply because all the powerful machinery (including excellent visualization tools) from (static) network theory~\cite{newman_book,barabasi_book} could be brought into action. In other words, for any discrete time step, one can understand and characterize the network using network theory and then couple the results for the sequence of times in some way. For this to make sense, the temporal resolution has to be low. Examples (mentioned above) when this could be fruitful include: ecological networks, interlinked habitats, brain networks and global trade networks.  

However, there are systems where graph sequences are not a good idea. Simply speaking, these are cases where the time resolution is high (or continuous) compared to whatever dynamic system on the network one would like to study, or when the time resolution is high and the contacts are instantaneous. E-mail metadata usually have a one-second time resolution, so a graph-sequence representation of an e-mail network (even a rather large one) would look like: empty graph, empty graph, empty graph, empty graph, graph with one link, empty graph, empty graph, etc.\ (Fig.~\ref{fig:lossless}(e)). If we study a disease spreading on a human proximity network, then, even if the graphs of the  sequence are non-trivial, it would probably be unreasonable to assume an infection could spread through more than one contact over a time step. In other words, the most fundamental assumption of static networks---that being indirectly connected (connected through a path of length larger than one) is different from not being connected---breaks down, so techniques from static network theory cannot be applied directly to
a graph of the sequence (or a layer of a multilayer network) Fig.~\ref{fig:lossless}(e). There are several ways proposed in the literature to get around this issue, but thinking of the graphs of a graph sequence as static networks could be misleading.

\subsubsection{Dynamic networks} \label{sec:dynamic}

One pillar of network science is the modeling of emergent properties of graphs. A canonical example is the Barab\'asi-Albert model~\cite{ba_model}, which shows how a microscopic mechanism in the network growth (preferential attachment) can lead to a macroscopic property of the network structure (a power-law degree distribution). Other models (e.g.\ Refs.~\cite{gg_dyn,krivitsky}) allow nodes and links to disappear too. We will call the idea of a system as a static network evolving in time a \textit{dynamic network} (we are well aware of the confusion of terminology---sometimes what we call temporal networks are called dynamic networks). Isn't this situation just the same as any temporal network? To some extent, the answer is like in the previous section---if the dynamic systems on the network are faster than the contact dynamics, and the network at any given moment of time is non-trivial, then yes. However, there is also a subtle difference of the research questions about them. For dynamic networks, the focus is typically on one class of networks (say protein interaction networks) and questions concern the structure of this network class, how the structure has evolved and how it affects dynamic systems on the network. Temporal networks are typically more data oriented---researchers investigate a data set, its structures, and how e.g.\ epidemic outbreaks would behave on it. Then one asks how these observations generalize by comparing results for different data sets. Perhaps, this slightly different approach comes from that there are no semi-universal structures in temporal networks that involve both the time and topology (as opposed to scale-free degree distributions in network theory~\cite{ba_model,barabasi_book} or bursty behavior of human activity~\cite{barabasi_bursts}).

\subsubsection{Time-node graphs} \label{sec:timenode}

Another way of representing temporal networks, akin to multilayer networks, is to make a network of time nodes (sometimes called the ``static expansion'' of a temporal network~\cite{michail_temporal_graphs})---representing the original node at a time. This type of network can be practical since it is straightforward to apply static network methods also over the time dimension~\cite{issue:taro,speidel_taro,pfitzner_betweenness}. Eventually one usually needs to map the time nodes back to the original nodes; maybe one exception could be a certain year's edition of an annual event (cf.\ ``NetSci 2015''---a conference during which parts of this paper was written).  (A time-node representation of our example network is shown in Fig.~\ref{fig:lossless}(d).)

\subsubsection{Time series of contacts on a static graph} \label{sec:static}

Contact sequences correspond, to some extent, to link lists in static network (a $2\times M$ matrix of the two nodes of every link in the network; where $M$ is the number of links). The other important scalable data structure for static graphs---adjacency lists (listing node-by-node all the neighbors of a node)---corresponds to assigning a time series of contacts to the links of a static graph. The advantage of thinking about temporal networks in this way is primarily visual---one can plot the underlying graph with all the powerful graph layout algorithms designed for static networks, one can even plot the time series of contacts as a time line. In practice, this only works for very small temporal networks, both because the underlying graph tend to be rather dense in empirical data, and because there is little space to visualize the contact time series (See Fig.~\ref{fig:lossless}(c) for an example.)

\subsubsection{Time-lines of contacts}  \label{sec:timeline}

Another, primarily visual, type of representation is a time line of contacts. Graphically, one would let one dimension represent time and one dimension the set of nodes. Then one connects two nodes by a line at the times of their contacts (Fig.~\ref{fig:lossless}(a)). The advantage with this representation is that the \textit{time-respecting paths} (sequences of contacts of increasing times) between nodes are very easy to identify---these are all paths that does not turn backwards in the time dimension. The disadvantage is, not surprisingly, that such a plot gets unintelligible if the temporal network has more than, say,  50 nodes. The visual information that survives longest is the temporal one, so with 50 nodes, one may be able to spot structure in the overall activity in the data, but not the structure of time-respecting paths.

One can also use time-lines of contacts between pairs of nodes, Fig.~\ref{fig:lossless}(b). This highlights the pairwise interactions, but the visual information relating to the topology, including the time-respecting paths is gone.

\subsubsection{Adjacency tensors}

Just like a static network can be represented as a binary matrix, an adjacency matrix, a temporal network can be represented as a binary tensor~\cite{gauvin_panisson,latent_gauvin,colizza_analytical,wehmuth,dunlavy_temp_linkpred,hamon_duality}. The pros and cons are also the same---an adjacency tensor, as a data structure, takes a lot of memory, but it allows for tensor algebra and all the neat and compact formulas that come with it. The memory problem can, in practice, be even bigger than for static networks since many empirical data sets (as mentioned) are very sparse in the time dimensions. It also shares some problems with the graph sequences and dynamic graph pictures. Since the dynamic system of interest may not be able to operate within the graph of a time step, the adjacency tensor cannot function like an (unnormalized) Markov transition matrix. Yet a complication is that time is directed, while many methods in tensor algebra assume indices could be relabeled (i.e.\ time order broken). Finally, we note a visualization method related to adjacency tensors---see Bach et al.~\cite{bach}.

\subsubsection{Film clips} \label{sec:film}

A natural way of thinking of temporal networks, especially visually, is to show them time step by time step, i.e.\ as a film clip~\cite{moody_viz,batagelj,grabowicz}. The obvious disadvantage is that one cannot see all the information at once. In developing methods, we feel it is a major problem not to be able to, for example, highlight a time-respecting path. To get a feeling for the overall activity and complexity of the data, however, it could be a useful approach.

\begin{figure}
\begin{center}
\includegraphics[width=0.8\linewidth]{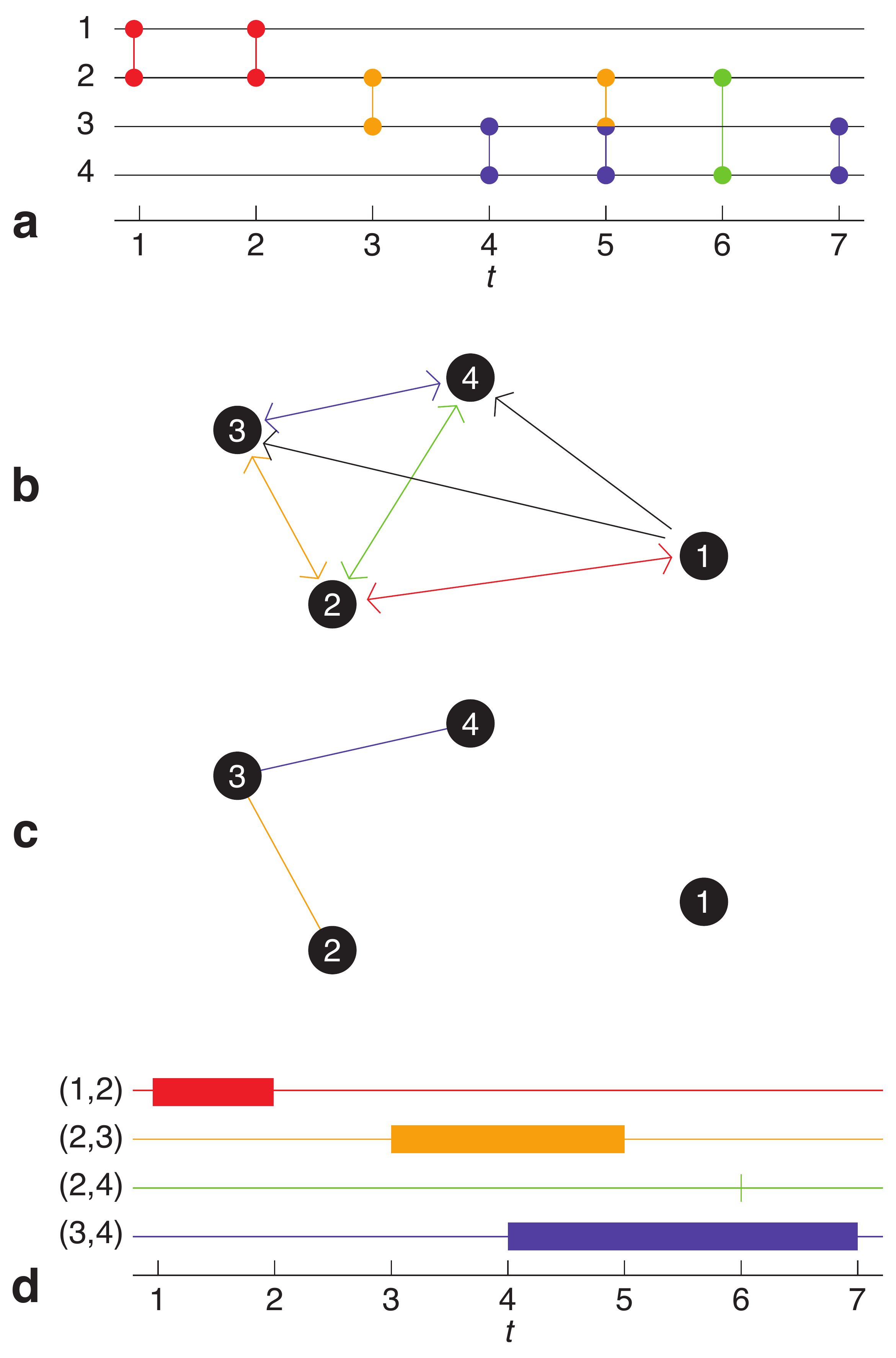}
\end{center}
\caption{Panel (a) shows an example contact sequence (the same as in Fig.~\ref{fig:lossless}). Panel (b) is the derived reachability graph at $t=0$. Panel (c) shows the time-windowed static graph from contacts between $t=3$ and $t=5$. Panel (d) illustrates a link-turnover graph where there is a link between two nodes if they had a contact before and will have one again.} 
\label{fig:lossy}
\end{figure}

\subsection{Lossy representations}

Now we turn to representations where some information of the original temporal network (typically contact sequence) is lost. We still talk about these as representations, not models, as the number of bits needed to encode them scale linearly with the system size.

\subsubsection{Weighted graphs}

A natural way of simplifying a contact sequence is to construct an link-weighted graph where the weight gives  a count of the number of contacts between two nodes. How useful such a representation is depends on what dynamic system one considers. Regarding random walks, Lambiotte et al.\ write: ``Contrary to standard procedures, the importance of an link is in general not proportional to its number of activations [\dots], but to the probability that it participates in the diffusive process.''~\cite{lambiotte_burstiness} Of course it might require simulations to calculate the probability of participation, and one might  have to settle for counting the number of contacts for that reason. The quality of a weight could be improved further if one knows when a dynamic phenomenon (e.g.\ a disease outbreak) starts. As an intermediate step to creating simple graphs, Holme~\cite{holme_comp_biol} uses weighted graph obtained by, for ever contact between a node pair, summing an exponential weight
\begin{equation} \label{eq:expweight}
\sum_i e^{-t_i/\tau}
\end{equation}
(where $t_i$ is the time between the beginning of the spreading and the $i$'th contact, and $\tau$ is a parameter that should match the time scale of the dynamic process).

\subsubsection{Reachability and influence graphs}

A reachability graph is a directed, unweighted graph that links $i$ to $j$ if there is a time-respecting path from $i$ to $j$~\cite{cheng_etal,moody,holme_saramaki_rev,batagelj_prapronik,nicozia_metrics,whitbeck}. Since the existence of time-respecting paths depends on time, such a graph needs to be a function of time. For many data sets that at any given time are very sparse (like email or phone-call networks) reachability graphs could be a good tool for visualization. The problem with reachability graphs is rather that they could be too dense---for our collection of temporal-network data sets (the ones used in Refs.~\cite{holme_info,holme_liljeros}), early in the sampling period, the reachability graphs have 30-100\% of the maximum number $N(N - 1)$ of directed links. In our example case, Fig.~\ref{fig:lossy}(a) the reachability network is very large at the beginning, see Fig.~\ref{fig:lossy}(b) with ten links present out of twelve possible. To remedy this some studies put a higher requirement on a directed link than just one time-respecting path, and defines directed ``influence networks''~\cite{taro_compsac,steeg_info}.

\subsubsection{Time-window graphs}\label{sec:window}

A simple way of reducing a temporal network to a static network is to include all the links present in a time window. (The very simplest way would be to take this time window as the entire sampling time.) Several papers address the question of choosing the time window so that the resulting graph is as useful and informative as possible. Krings et al.~\cite{bernhardsson} make a comprehensive study of this question with respect to mobile call networks. Psorakis et al.~\cite{psorakis} discuss this question applied to animal proximity networks. They show the static networks are dominated by clusters at short time scales and are more like models of heterogeneous graphs (cf.\ Ref.~\cite{ba_model}) for longer time scales. For short time scales they find the starting time also can make a big difference because of periodic patterns. In a study of proximity data, Sekara, Stopczynski and Lehmann~\cite{vedran_arek_sune} takes this idea a step further and uses the clusters that fall out from shortening the time windows to define a ``gathering''. They argue that gatherings are natural building block to characterize this type of data and the face-to-face interactions they measure. Walker et al.~\cite{walker_cdiff} discuss a similar characterization to understand disease spreading in hospitals. The time window size also affects modeling of dynamical systems on the network, Liljeros, Giesecke and Holme~\cite{liljeros_giesecke_holme} discuss how to chose the parameter values to simulate disease spreading on a time-windowed proximity data, in particular how the parameter values of the disease spreading model are related to the time window size. Ref.~\cite{holme_comp_biol} shows that the beginning of the time window should be as close as possible to the beginning of the outbreak for the reduced, static network to be as useful as possible for epidemic modelling. G\'enois et al.~\cite{genois_vacc} investigate the optimal time window of proximity data ahead of targeted disease-intervention effort.

\subsubsection{Concurrency graphs}

A classic theory from the network epidemiological literature~\cite{watts_may} explains that sexually transmitted diseases spread much faster and further if there are many concurrent partnerships in the population. Usually, ``concurrency''---as a network property---is thought to be a property of an entire graph \cite{kretzschmar_etal}. It is, however, implicitly clear from Refs.~\cite{kretzschmar_etal,morris_kretzschmar_concurrent}, that one can define the concurrency of a person as the number of ongoing partnerships at some specific time $t$, i.e.\ the degree in a graph where an link means that a contact has happened before $t$, and will happen again. A bit more generally, Ref.~\cite{holme_comp_biol} defines links in concurrency graphs as pairs of nodes having contacts both before $t_\mathrm{start}$ and after $t_\mathrm{stop}$.

\subsubsection{Difference graphs}

Neiger et al.~\cite{neiger} use a static graph highlighting change (rather than persistent patterns, like the above concurrency graphs). They define a \textit{difference graph} between two consecutive time steps by the links that changed between the time steps.

\subsubsection{Memory networks}

Memory networks is a representation of a slightly more informative data sets than contact sequences. It assumes there is a recorded walk process on a set of nodes, like flight passengers on multi-hop itineraries. This could naturally be simplified to a temporal network of moves where one move is independent of the others, which could be further simplified to a weighted network (effectively a Markov process transition matrix). Rosvall et al.~\cite{rosvall_memory,lambiotte_salnikov} argue that one can encode much information without much more complexity by representing this data as a second order Markov process, i.e.\ one records the frequency of walkers arriving to node $i$ from $h$ that continues to $j$.

\subsubsection{Static graph with a model for the contact dynamics} \label{sec:ongoing}

Above, we have considered simplified representations of temporal networks that project out the temporal dimension (or change the meaning of the time dependence, so that paths of links could represent nodes being indirectly connected). One can also simplify a temporal network in ways that retain some of the temporal structures. One such way is to separate the topology and the contact dynamics (sometimes~\cite{speidel_rw} called \emph{link activation}) over links. The general idea of thinking of a temporal network as a static network with some process generating contacts over the links is called the \emph{ongoing link picture} in Ref.~\cite{holme_liljeros}. The very simplest way would be to assume an underlying static network where the contacts between the links are modeled by the same interevent time distribution for all the links. Goh, Vazquez, et al.~\cite{min_goh_vazquez,thebook:vazquez} use such a set-up to argue that fat-tailed interevent times slow down spreading phenomena.

Inferring this type of picture from data is problematic, even when it correctly describes the data. One thing missing is the beginning and end of links. For a short sampling time window, this will, in principle, be a smaller problem (since the turnover of relationships is a slower process than the contact, or link-activation, dynamics). So in that sense, smaller sampling times are better for this picture. However, for a short sampling time, the chance a link will not have a single contact, and thus being unobserved, is larger. There are ways to infer the beginning and end times, and how good the assumption that there is no turnover of links is~\cite{kivela_porter,holme_ongoing,holme_comp_biol}. One could also use link prediction~\cite{lu_zhou} to infer missing links, such methods are (to our knowledge) not capable of estimating how many missing links there are.

\subsubsection{Birth and death (but not rebirth) of links} \label{sec:linkturnover}

Another way of simplifying a temporal network, that keeps both some topological and temporal information is to consider a link continuously active between its first and last contact. If the simplified picture of the previous section is more accurate for short sampling interval, this picture (called \emph{link-turnover picture} in Ref.~\cite{holme_liljeros})  is better for long sampling durations---after all, regardless the system in question, all links needs to at least have a beginning. This is similar to dynamic networks (Sec.\ Ref.~\ref{sec:dynamic}), only that it does not allow links to reappear. If the nodes, perhaps by virtue of their links, also enter and leave the system, then spreading processes have a predominant direction highlighted by this picture---from the older to the newer nodes. Holme and Liljeros~\cite{holme_liljeros} argue that for disease spreading on many of the available data sets, the link turnover picture is more accurate than the idea of a static graph with an overlying contact dynamics. Note the connection between the link-turnover picture and concurrency graphs discussed above---the active links at time $t$ in the link-turnover picture defines the concurrency graph at $t$.

\section{Temporal network structure}

At the heart of network science lies the idea that there is information in the structure of a network that can tell us something about both the network evolution and systems that operate on the network. Indeed, a definition of ``structure'' in this respect, could be ``what carries information in a temporal network''.  In this section, we will discuss how to measure the structure of temporal networks.

\subsection{Network topology}

This colloquium concerns temporal networks, and by ``topology'' one usually refers to static structures. There are plenty of literature about this (see e.g.\ Refs.~\cite{newman_book,barabasi_book} and references therein). Here we will discuss the  role of network topology measures in  temporal networks.

In the history of static network theory, measuring network structure have been  driving the field. For example, after Barab\'asi and coworkers discovered how common scale-free (i.e.\ power-law-like) degree distributions are~\cite{ba_model,barabasi_book}, there was a huge effort both to measure degree-distribution and to model their emergence. For temporal networks, similar ubiquitous structures are yet to be discovered, perhaps they do not even exist. This has led the research in temporal networks down a slightly different path, where the focus is more on dynamic systems on the network and how they are affected by structure, and less on discovering common patterns or classifying networks.

Nevertheless, many temporal network studies are more or less motivated by static network theory. Many authors try to generalize a network-topology measure (we will see plenty of examples of this in the sections on distance metrics, centrality measures and community structure methods below). Another way static network theory enters tem\-po\-ral-net\-work  papers is through temporal mechanisms that gives static network of accumulated contacts their structure. Mantzaris and Higham~\cite{thebook:higham1} investigate triadic closure---a mechanism behind e.g.\  the high clustering coefficients of social networks~\cite{newman_book} (i.e.\ a high density of triangles). Kunegis et al.~\cite{kunegis} study aspects of the preferential attachment mechanism that can give static networks an emergent power-law degree distribution. Some studies also try to separate topological and temporal effects on spreading phenomena~\cite{holme_masuda} or diffusion~\cite{luis} (more about that below).

\subsection{Temporal structure}

\subsubsection{Burstiness and interevent time statistics}\label{sec:burst}

Now we turn to structures that only concerns temporal aspects. Just like topological measures, these cover only one aspect of temporal network structures. A very common temporal structure in the literature is the interevent time distribution. In a time series of events this is the frequency distribution of the time between the events. If the events are independent and drawn from a uniform distribution, then the interevent time distribution will be exponential. In empirical data sets, however, it is usually fat-tailed, or even scale-free~\cite{holme_ongoing,johansen_resptimes,barabasi_bursts}. A difference to scale-free degree distributions is that bursty time series are usually characterized by their coefficient of variation (called ``burstiness''~\cite{goh_barabasi,thebook:goh}) rather than their power-law exponent.

In a temporal network, one could think of a few different types of interevent times, corresponding to a few different types of time series. The most basic is the time series of contacts between two nodes. The problem with  burstiness of individual links is that there are usually too few data points to measure it accurately (or to measure it at all). In many types of data sets, the number of contacts per links also has a fat-tailed distribution~\cite{holme_ongoing}. This means that it will be problematic to average the link burstiness, since most links have only a few contacts and thus a poor estimate of burstiness. Some authors~\cite{karimi_ramenzoni} concatenate the interevent times and measure the burstiness of that time series, but this does not get around the problem that individual nodes and links follow individual patterns. 

Next, one can measure the burstiness of individual nodes. This makes most sense when the temporal network records a distinct sender and receiver of a contact. The bursty structure of when people send e-mails~\cite{eckmann} was an important discovery for this line of research. The distribution of when people receive e-mails is perhaps less interesting (and less bursty for that matter), but not without structure (like daily patterns)~\cite{holme_ongoing,jo2012circadian,tala0,tala}. The connection between the burstiness of links and nodes were investigated in Refs.~\cite{karsai_bursts1,karsai_bursts2}.  

\subsubsection{Time aspects of network evolution} \label{sec:timeas}

In addition to interevent times, there are many other purely temporal structures. For example, the time between the first and the last contact (between two persons or of one person) in a data set~\cite{holme_ongoing,thebook:miritello,miritello_limited}. This could be taken as the life time of a link or node in the data (how to compensate for the sampling bias from a finite sampling time was discussed in Refs.~\cite{holme_ongoing,kivela_porter}).

Finally, at the largest scale, one can measure the growth or decay of a temporal network. In some data sets, all the nodes and links are basically present at the beginning of the sampling, and stay to the end (this is the case for several cell-phone data sets~\cite{jari_moro}, e-mail data~\cite{eckmann} and proximity data~\cite{pfitzner_betweenness,conference}). In other data sets, the overall activity, including the number of nodes and links, grows (this is the case for the online prostitution data studied in Ref.~\cite{rocha_liljeros_holme} and the online communities~\cite{holme_edling_liljeros,karimi_ramenzoni}). In yet other data sets, there is a constant turnover of nodes~\cite{gallery}. One metrics capturing the presence of  nodes and links throughout the data is the fraction of them that are seen both before time $fT$ (from the start of the sampling) and after $(1-f)T$, where $T$ is the sampling duration. Holme and Masuda~\cite{holme_masuda} use $f=0.05$ and studies time not only as the real time but also as the number of a contact, counting from the start (cf.\ Refs.~\cite{albano_ex_intrinsic,albano_time_scales,issue:saramaki_holme}). Another measure for a similar purpose---to monitor the growth of the network---is to measure the fraction of nodes or links present at half of the sampling time, or half of the total number of contacts~\cite{holme_masuda}.

\subsection{Paths and generalized distances}

The most basic quantity of static networks that explains the relationship of two nodes is their distance, i.e.\ the smallest number of links of a path connecting them. In temporal networks there are many ways of generalizing distance and which one to chose is not always obvious. A classic idea is to consider nodes $i$ and $j$ at time $t$, then the \textit{latency}~\cite{lamport} is $t$ minus the latest time when there is a time-respecting path from $i$ to $j$. One could also look forward and define the \textit{temporal distance}~\cite{pan_saramaki} $\tau(i,j,t)$ as the earliest time to reach from $i$ to $j$ on a time-respecting path starting at $t$. The term ``distance'' may, at first, feel awkward as a quantity of the dimension time, but one way of thinking of the temporal network is as an operator that sets the speed of processes, i.e.\ converts space to time, and vice versa.

Rather than latency and temporal distance, one would many times want to have a time-independent generalized distance. The problem is that for any empirical data set, for late enough $t$, the temporal distances are infinity (or, analogously, for early enough $t$, the latencies are infinity). Pan and Saram\"aki~\cite{pan_saramaki} explore different options---either to assume periodic boundary conditions or separate the issues of whether a node can be reached and how long time it takes to reach it given that it can be reached. The latter approach was also proposed by Holme~\cite{holme_reachability}, who defines \textit{reachability time} as the average shortest time to go from $i$ to $j$ over the times when there is a time-respecting path between them.

Another generalization of distances to temporal networks is to consider the number of contacts in a temporal path. Refs.~\cite{bayhan_hoplimit,batagelj_prapronik} use only this definition, while e.g.\ Refs.~\cite{buixuan_etal,huanhuan_path,casteigts,starnini2012} use both.

Computing, generalized distances in time from one node to all others is rather straightforward---just simulate an SI spreading with 100\% transmission probability and keep track of the time  when a node gets infected, that will also be the shortest value. For hop-counts, one could use a Ford--Fulkerson type algorithm where one runs through the contacts in time order and the currently shortest path is updated for each node involved in a contact. Proofs and technical details about such algorithms can be found in Refs.~\cite{lamport,buixuan_etal,huanhuan_path}. See Ref.~\cite{iyer} for some advanced algorithmic issues about estimating temporal-path based metrics in streaming data. Finally, we note that the terminology is exceptionally confusing regarding the topic of this section---there are a handful of terms for all the mentioned concepts, and some terms are used differently by different authors.

\subsection{Centrality measures}

Centrality measures are usually comparatively easy to adapt from static networks. In traditional network theory~\cite{newman_book,barabasi_book} they typically build either on some assumptions of a (more or less abstract) dynamic system on the network, or on graph distance. One can simply let the dynamic system evolve following the contacts rather than the links or replace distance by latency or temporal distance. The only major difference is that a centrality measures will depend on time  (just like distance metrics). Alternatively, one could project out time either by taking an average or the minimum. Closeness $C_C(i,t)$, for example, that for static networks are defined as
\begin{equation}\label{eq_cc}
C_C(i)=\left[\frac{1}{N-1}\sum_{j\neq i} d(i,j)\right]^{-1}
\end{equation}
where $d(i,j)$ is the graph distance. This can straightforwardly be extended to temporal networks as~\cite{huanhuan_path,nicozia_metrics,batagelj_prapronik,hyoungshick}
\begin{equation}\label{eq_cc_t}
C_C(i,t)=\left[\frac{1}{N-1}\sum_{j\neq i} \tau(i,j,t)\right]^{-1}~.
\end{equation}
A problem with closeness centrality in static networks which becomes much aggravated for temporal networks is that there might not be any path between $i$ and some of the other nodes. A workaround is to average the inverse distance rather than inverting the average distance~\cite{pan_saramaki}
\begin{equation}\label{eq_cc_t_mod}
C_C(i,t)=\frac{1}{N-1}\sum_{j\neq i} \frac{1}{\tau(i,j,t)}~.
\end{equation}
The problem with this approach is that this type of mean is less intuitive as it is a somewhat arbitrary combination statistic of the size of the components~\cite{nicosia_components} and the (temporal) distance within them. One could actually also ignore the distances---the up and downstream components of a node are  also a sort of centrality measures. Kim et al.~\cite{mascolo_cent_pred} discuss how to estimate static centrality metrics of a time-window aggregated network (cf.\ Sec.~\ref{sec:window}) from a previous time window. Michalski et al.~\cite{michalski} discuss an interesting way of omitting the time dependence by weighing older paths lighter. One can define temporal betweenness centrality~\cite{newman_book} in a similar way~\cite{berger_wolf_tbetweenness,nicozia_metrics,michalski,hyoungshick,thebook:latora,ser_giacomi,AhHi15}. Takaguchi et al.~\cite{issue:taro} define ``temporal coverage centrality'' of $i$ as the fraction of node pairs $(j,j')$ such that passing  $i$ would not increase the time to reach from $j$ to $j'$. Williams and Musolesi~\cite{williams_musolesi} define path-distance-based centrality measures for spatiotemporal networks (where one also knows where the nodes are). Takaguchi et al.~\cite{taro_individual} consider a different notion of path-based centrality, or importance---not of nodes but of contacts. Simply speaking, a contact is defined as important if it speeds up many time-respecting paths.

Rocha et al.~\cite{rocha_rw_cent} use the occupation probability of random walks on a temporal network as a centrality measure. The ``communicability'' metrics of Estrada et al.~\cite{communicability_evolving,estrada_communicability,thebook:higham2} is also based on a random walk through a quantum theory propagator. Some related measures were also scrutinized by Rogers~\cite{rogers}. Random walks centralities overlap with centrality measures defined through spectral properties of the adjacency matrix (or related matrices). Praprotnik and Batagelj~\cite{batagelj_centrality} make extensions of matrix-based centrality measures for static networks (like eigenvector and Katz centralities) to temporal networks. Taylor et al.~\cite{eigentaylor} define such a method that overcomes some obvious problems like how to project out the time in a principled and meaningful way. 
A different approach to centrality was taken by Pan and Li~\cite{li_controllability} who define ``control centrality'' roughly speaking a node's ability to control the network (in a control theory sense). Zhang and Li~\cite{li_c} and Ghoshal and Holme~\cite{ghoshal_holme} define centrality-like measures capturing how much an individual participates in the activity of a temporal network. Grindrod and Higham~\cite{GHrsoc} develop a differential equation based centrality measure which generalizes Katz centrality~\cite{newman_book} of static networks. In another paper~\cite{GHsr}, Grindrod and Higham propose a path-based centrality measure that downweigthts long and old paths. In yet another paper~\cite{LMGAOH13} these (and some other) authors validate centrality measures for a temporal Twitter network by (among other ways) a panel of experts.

A problem related to centrality is to rank players and teams in competitive sports. Motegi and Masuda~\cite{motegi_masuda} address this issue in a situation when one have an incomplete time-annotated set of matches with a win-or-lose outcome. More specifically, one would like to rank the teams or players so that there are as few upsets (lower ranked player beats a higher ranked player) as possible. At the same time, newer results should be more important than newer.

\subsection{Controllability}

Structural controllability is a concept that has been adapted from static network theory~\cite{controllability} to temporal networks. It assumes a system with in- and output terminals connected into a network. The dynamics between the nodes is assumed to be simple, so that the output from one node is proportional to (or at least a monotonous function of)  the input. One assumes no time delays in the dynamics and no more complex effects (like memory etc.). One can show that, for static networks, there are very simple topological characterizations of what nodes that one needs to control in order to control the entire network. P\'osfai and H\"ovel~\cite{posfai_structural} and Cimatti et al.~\cite{cimatti} present comprehensive generalizations of this theory to temporal networks. P\'osfai and H\"ovel~\cite{posfai_phasetransition} also show that for some classes of temporal networks, there can be a phase transition in the time scale of the dynamics between a state where the network is controllable by a a vanishing faction of nodes, and a state where a finite fraction needs to be controlled. Pan and Li~\cite{li_controllability} make an equally ambitious study of, among other things, limits of the number of nodes that need to be controlled. In another paper, Pan and Li~\cite{li_towards} discuss a graphic characterization of the nodes controlling a temporal network.

\subsection{Other graph invariants}

For static networks there were many early studies  characterizing properties of the network structure by a single number. Functions that map a graph (regardless of the labeling of the nodes) to one number is called a \textit{graph invariant}. The number of nodes or links, are two simple examples. In static network theory some other graph invariants are the clustering coefficient (measuring the fraction of triangles) and the assortativity (measuring the tendency of nodes of similar degrees to attach to each other)~\cite{newman_book}. For temporal networks, we already mentioned some trivial graph invariants in Secs.~\ref{sec:burst} and \ref{sec:timeas}. All in all, there are not so many temporal network studies that proposes new functions to characterize the joint structure of time and topology. One example is the average number of nodes reachable from a random node at a random time (within the sampling time) called ``reachability'' in Ref.~\cite{holme_reachability}. Another example is Pfitzner et al.'s ``betweenness preference'' study~\cite{pfitzner_betweenness}. They investigate the predictability of time-respecting paths. First they define a matrix, for every node $i$ and time step $t$, saying for which pairs of nodes is $i$ situated on a time-respecting path between them. Through a series of manipulations, including some information theory, the authors arrive at a measure capturing the tendency for paths to be time-respecting. Next, we mention Scellato et al.~\cite{scellato_robu} who define a measure of a temporal network's robustness as the expected change in the average closeness centrality (Eq.~\ref{eq_cc_t_mod}) if a random node is deleted. Finally, a way of summarizing the activity of a node in a time window using ``factorial moments'' was proposed by Chi and Yang~\cite{liping}. By this method (adopted from high-energy physics) one can obtain scaling exponents characterizing the fluctuations of node or link activity as a function of the duration of the window.

\subsection{Entropy measures}

Information theoretical measures have been increasingly popular the last few years, not only in temporal network theory. Entropy measures capture the randomness of e.g.\ the contacts in a temporal network. If the entropy is low, there is much regularity in a signal which also means that it is easy to predict. For example, Takaguchi et al.~\cite{taro_prx} use an  entropy-based analysis to conclude that knowing a current face-to-face conversation partner decreases the uncertainty of who the next conversation partner will be  by about 30\%. Kun et al.~\cite{thebook:bianconi} use entropy to understand how to estimate the probabilities of configurations of face-to-face interaction. Perotti et al.~\cite{sparsity} use a entropy based metrics ``sparsity'' to characterize events in temporal networks. Their intention is to create a metrics orthogonal to burstiness and other interevent time statistics by first looking across nodes in a time window, then projecting out the time dimension. Some studies use entropy rates of random walks on temporal networks to characterize the temporal network structure~\cite{issue:saramaki_holme,scholtes_causality,rosvall_memory}.

\subsection{Persistent patterns} \label{sec:persistent}

Obvious patterns to look for in temporal networks are links and subgraphs that do not change as much as others. Clauset and Eagle~\cite{clauset_eagle} measure the auto correlation function of links in temporal networks. Valdano et al.~\cite{valdano_poletto} introduce a similar metric but for nodes. They define ``loyalty'' as the Jaccard index between the neighborhoods of a node at one time step and the previous time step, and show that it is important to understand the role of nodes in disease dynamics. Zhang et al.~\cite{xin_motif} also argue that loyalty in socio-economic systems leads to heavy tails of the duration of business contacts. Neiger et al.~\cite{neiger} address an inverse problem by measuring how connected changing links are. Briefly, they construct a network of the links that changed from one time step to another, and measure the minimum node cover (set smallest set  of nodes such that each link of the graph is incident to at least one node of the set) of this network.

\subsection{Cyclic patterns}

In empirical temporal networks, especially those related to human activity, there will be cyclic patterns. Several authors~\cite{jo2012circadian,tala0,tala,holme_ongoing} discuss  aspects of how to handle this type of patterns. Both how to measure how strong such tendencies are, and how one can get rid of them (for analysis of quantities where a changing background activity level is undesirable). Lahiri and Berger-Wolf~\cite{lahiri_berger_wolf_2008} propose an axiomatic formalism to handle such phenomena. Their method is flexible enough to allow some nodes and links missing from a period and still regard a subnetwork cyclic. On the other hand, their method does not handle small random shifts in the cycle length, so one would need to first coarse grain the data to be able to observe almost cyclic pattern.

\subsection{Motifs}

A \textit{motif} in a static network is a small subgraph that is overrepresented in a graph compared to in a null model (typically random graphs constrained to have the same number of nodes, links and degree sequences as the original graph). One could imagine several ways to extend this concept to temporal networks. Zhao et al.~\cite{zhao_etal} and Kovanen et al.~\cite{kovanen_motif} focus on contacts that connect a group of individuals and all happen within a time window of size $\Delta t$. They also study different classes of such motifs. Using this idea both  papers examine mobile phone data sets. Kovanen et al.\ discover gender-specific patterns and homophily (the tendency of similar individuals to be connected). These ideas are explored in greater detail in Ref.~\cite{thebook:kovanen}). Zhao et al.\ argue that complex motifs, like ping-pong patterns are very overrepresented. Just like triangles are common in static social networks, temporal patterns involving the links of a triangle within a short time period are common and important for spreading phenomena~\cite{thebook:higham1,li_clustering,issue:saramaki_holme}. Rocha and Blondel~\cite{rocha_flow} also use counts of triangles to characterize networks. Hulovatyy et al.~\cite{hulovatyy} extend the frameworks of Refs.\cite{kovanen_motif,zhao_etal}. First, they define a more restricted class of temporal subnetworks---$\Delta t$-causal subgraphs (those are such that you can reach between all pairs of participating nodes by time-respecting paths. Second they define ``temporal graphlets'' as the equivalence classes, with respect to the order of events, of $\Delta t$-causal subgraphs, and argue it is useful to think of them as building blocks of temporal networks. In Li et al.~\cite{mantegna} the authors consider the order in which static network motifs  are assembled---an extension of the network motif concept that is a bit closer to its original.  Zhang et al.~\cite{xin_motif} define temporal motifs between to consecutive time steps in a bipartite network. Lahiri and Berger-Wolf discuss how to predict recurring static motifs in Ref.~\cite{lahiri_berger_wolf_2007}. Redmond and Cunningham~\cite{redmond_cunningham} investigate the related algorithmic problem of counting and listing isomorphic temporal subgraphs.
 
\subsection{Mesoscale structures} \label{sec:meso}

Network motifs are a way to look at how groups build up a temporal network from the small scale. It rests on the assumption that not everything in the network is equally interesting. Another, even more popular, approach does not make this assumption, but rather asks how a network can be divided into groups so that all nodes belongs to one group (the relative shape of the groups do not matter). This approach is to find so called \textit{mesoscale structures} (a phrase borrowed from the natural sciences, but not used completely analogously). The most common mescoscale structure is \textit{community structure}~\cite{fortunato} and that is what we will focus on in this section. We note, however, that recent years have seen an increasing interest in core-periphery structure~\cite{mejn_cp,porter_cp}. 

In a static network, a community is thought of as a cohesive subnetwork---more densely connected within than to other subnetworks. To operationalize such a definition is a very open task---it seems almost like for every concrete decision you have to make in creating a community detection algorithm there is no obviously best choice. Therefore there is a huge number of community detection methods~\cite{fortunato}. For temporal networks, the easiest approach is of course to separate the time and network dimensions. In other words, at time $t$ one would first run the community detection algorithm on the static network to decompose the nodes $V$ into communities $c_{1{,}t},\dots,c_{n(t){,}t}$ so that $\bigcup_i c_{i{,}t}=V$. The next step is to merge communities at $t$ with overlapping communities at $t-1$. There are many ideas in the literature how to do that.  Ref.~\cite{seceder} maps the indices of $t$ to the indices of $t-1$ so that the sum of mismatching indices is minimized. This approach is very simplistic and some pathologies---if the groups are $\{1,\dots,50\},\{51,\dots,100\}$ at $t-1$ and $\{1,\dots,99\},\{100\}$ at $t$, then the algorithm would consider $\{51,\dots,100\}$ leaving a group and joining another, while it would be more natural to think of the two groups as merging and a single-node group forming. Sekara et al.~\cite{vedran_arek_sune} merge communities (or, rather, ``gatherings'', see Sec.~\ref{sec:window}) by hierarchical clustering. In other words, they construct a hierarchical order (dendrogram) of the communities by first assigning a pairwise coupling strength
\begin{equation} \label{eq:overlap}
1-\frac{|c_t\cap c_{t'}|}{|c_t\cup c_{t'}|}e^{-\gamma(|t-t'|-1)} ,
\end{equation}
where $c_t$ and $c_{t'}$ are two communities at two different times $t$ and $t'$.
    The idea of the exponential factor is to decrease the  weight between groups the further apart they are in time. Sekara et al.~\cite{vedran_arek_sune} further discuss how to break the dendrogram into temporal communities.
Tantipathananandh et al.~\cite{tantipathananandh_etal} propose a much more elaborate scheme which penalizes a node for being temporarily outside of a group, or  changing a group, or starting a new group, with a cost. Then the temporal group dynamics is inferred from minimizing the total cost.
Kauffmann et al.~\cite{kauffmann} also design a cost-based time-clustering scheme. Folino and Pizzuti and the early paper by Mucha et al.~\cite{mucha_etal} use an approach common in static networks---to maximize an objective function (a.k.a.\ \emph{modularity function}). Pietil\"{a}nen and Diot~\cite{pietilanen} use modularity maximization and subsequent aggregation in the time dimension. He and Chen~\cite{he} recalculate the partitioning from a previous time step and thus saves time compared to calculating it from scratch every time step. 

Rosvall and Bergstrom~\cite{rosvall_alluvial}, Bazzi et al.~\cite{bazzi} and Chen et al.~\cite{chen_kawadia} present other methods where the clustering in clustering in time comes from the persistence of clusters of time-sliced networks. Gauvin et al.~\cite{gauvin_panisson} use a tensor factorization approach to identify additive temporal subnetworks. They validate the method by recreating the class structure of a face-to-face interaction network of a school~\cite{school}. Matias and Miele~\cite{matias_block} use a dynamic stochastic block model to find temporal communities. Cai et al.~\cite{cai} propose a measure to characterize the activity level of communities. Peixoto~\cite{peixoto} discusses a stochastic blockmodel approach that puts boundaries to the groups in time such that it is reasonable to ignore the precise timing of contacts within a time window (just like one can ignore the wiring of a community in a static stochastic block model). Peel and Clauset~\cite{peel} take a similar approach in their identification of change points in temporal networks. Speidel et al.~\cite{speidel_taro} define a community detection scheme on time-node graphs (cf.\ Sec.~\ref{sec:timenode}).

Given how different temporal networks are (when it comes to path statistics and spreading phenomena), we can envision mesoscale structures that are further from static networks than the ones mentioned above. For a start it would be interesting with a community detection method that does not need  aggregation over time windows (for the sparsest temporal networks, say a email network at a second resolution); that e.g.\ operates directly on the flow of some dynamic system.

\subsection{Time scales}

Timescales is a concept that temporal-networks researchers frequently use. The basic type of reasoning with time scales are of course common knowledge. The dominant contaminating isotope after the Chernobyl disaster was Caesium-137. It decays into non-radioactive isotopes exponentially, with a time constant of 44 years. Thus (ignoring dispersion by wind and rain) radiation problems should vanish in a time scale of centuries. With temporal networks, however, this type of reasoning becomes, as we argue in this section, more complex.

A typical definition of temporal networks, opposed to evolving static networks, is that the time scales of the network evolution are of the same order, or shorter, than dynamic processes (epidemic outbreaks, Internet traffic, etc.) on the network. If the time scale of the evolution is longer than the dynamic processes, then one can assume the dynamic process is finished by the time there is a change to the network. However, for many reasons, it is rarely that simple. First, in empirical temporal networks, there are many types of processes that all can take place at their own time scale. Take a proximity network of a researcher and her colleagues as an example. Passing a colleague in the corridor could take seconds, a chat could take minutes; a meeting hours; a conference days; a research project months; the entire acquaintanceship could take years or decades. All of these processes could leave an imprint on the temporal proximity network. These aspects (with application to mobile phone interaction) are discussed further in Ref.~\cite{jari_moro}. Second, the nodes could follow very different individual patterns, and that could have a big effect on any kind of dynamics on the network. This is known to be an important factor for the spreading of sexually transmitted infections~\cite{liljeros_edling_amaral} and information in social media~\cite{holme_ieee}. Some papers~\cite{clauset_eagle} find time scales for individual links. Third, it could be the case that some types of interactions are described by a scale-free temporal statistics. This means that there is no well-defined time scale. The most well-studied case of this category is power-law distributed interevent times~\cite{min_goh_vazquez,barabasi_bursts}. If the exponent $\gamma$ of the power-law is large enough, the distribution would have a well-defined average, but if $\gamma<1$, it does not, meaning if we monitor the average of the distribution in an ongoing measurement, there would always be an observation long enough to change the average. Even if one can characterize narrower power-law distributions (and other fat-tailed distributions) with mean, median, variance, etc., they do not define a scale in the sense of the time constant of an exponential decay. Four, the accuracy of the time-scale measurements can vary much, and one may be tempted to include the more accurate ones (e.g.\ diurnal patterns) in the modeling even though the less accurately measured ones are more important for dynamic systems. Five, some relationships (e.g.\ interevent time or link lifetime distributions) cannot be easily parametrized by quantities  of the dimension time, thus defining time scales. This is  akin to a model selection problem, where the number of time scales one can identify comes from a trade-off between the goodness of fit and the simplicity of the model.
The sixth reason that time scales are troublesome in empirical data is that they can be hard to separate from effects of the dynamic system itself. This is a more elusive problem and perhaps rarely very severe. If one studies e.g.\ rumor propagation in social media, and imagine this spreading happens on a temporal network of  follower-type contacts, then it could be that who follows whom depends on the information spreading, so that eventually time scales of the spreading dynamics could mingle in with time scales of the activity of the social media.

Caveats aside, it is clear that time scales are so useful that it makes no sense to simply ignore them. There are also several attempts to define and analyze time scales. One common approach comes from temporal community detection. As discussed (Sec.~\ref{sec:meso}), it is natural to think of temporal network communities as subgraphs that are densely connected within and sparsely connected to other communities during a time window. The choice of the interval should be such that the community is relatively stable throughout the interval. If the same process governs the formation and  dispersion of communities in the entire system, one could find a common time scale for the process by temporal community detection. Several authors have implemented this idea~\cite{delvenne_timescales,tantipathananandh_etal}. 

Another approach to defining time scales is to look at the behavior of a dynamic system on the network. Different aspects of the temporal network structure could affect different dynamic systems, so this method cannot map out all kinds of time scales in a data set. For this purpose, researchers have used both random walks~\cite{baronchelli_resolution} and spreading processes~\cite{bayhan_hoplimit}.

Caceres and Berger-Wolf~\cite{thebook:bergerwolf} define timescales from the optimal time windows. ``Optimal'' here could be evaluated in several different ways. It could, for example refer to how compressible (in an information theoretic sense) the temporal network within the time window is (which should reflect how regular, e.g.\ persistent, the activity is within, cf.~\cite{rosvall_alluvial}). Within this framework, Fish and Caceres~\cite{fish_caceres} define time scales from the optimal time window for link prediction.

Yet a different approach to time scales comes from Ref.~\cite{lentz_selhorst} who think of processes taking place on the nodes, so that a ``dilution of the temporal network occurs, when the intrinsic node time scale (node waiting time) is much larger, than the edge dynamic (temporal resolution).''

\section{Manipulating, predicting and generating temporal networks}

In this section, we investigate a number of way to analyze temporal networks, either by manipulating an empirical data set or simulating a dynamic system on the temporal network.

\subsection{Randomization and similar reference models} \label{sec:random}

The idea of randomization techniques is to understand the effect of a structure by destroying it through randomization. If one measures e.g.\ the speed of a spreading process in an empirical temporal network and  an ensemble of temporal networks where this particular structure is randomized, then one can see how much faster or slower the spreading becomes because of this structure.

Randomization techniques are much more powerful for temporal networks than for static networks. First, they are much more versatile. For static networks, they are basically restricted to shuffling links (keeping the degrees of the nodes fixed), arriving to something similar to the configuration model with the same degree sequence as the empirical network~\cite{newman_book}. For temporal networks, we will see a multitude of randomization schemes (and still not mention all conceivable---see Refs.~\cite{holme_reachability,holme_saramaki_rev,holme_ieee} for more). The main reason, however, that they are more powerful for temporal networks is that one can study effects of correlations in the real data set without having to make an exhaustive list of the correlations. In static networks, it is usually more fruitful to build in the correlations into the model from scratch (in e.g.\ some extended configuration model~\cite{newman_book}). If one take that approach in temporal networks, there are simply so many types of structures that it is hard to say if an observed result comes from the generated correlation, or from some other structure that was not generated. Another advantage is that, with  randomization analyses one can be arbitrarily close to the real data, and investigate the structure small steps away.

\begin{figure}
\begin{center}
\includegraphics[width=0.8\linewidth]{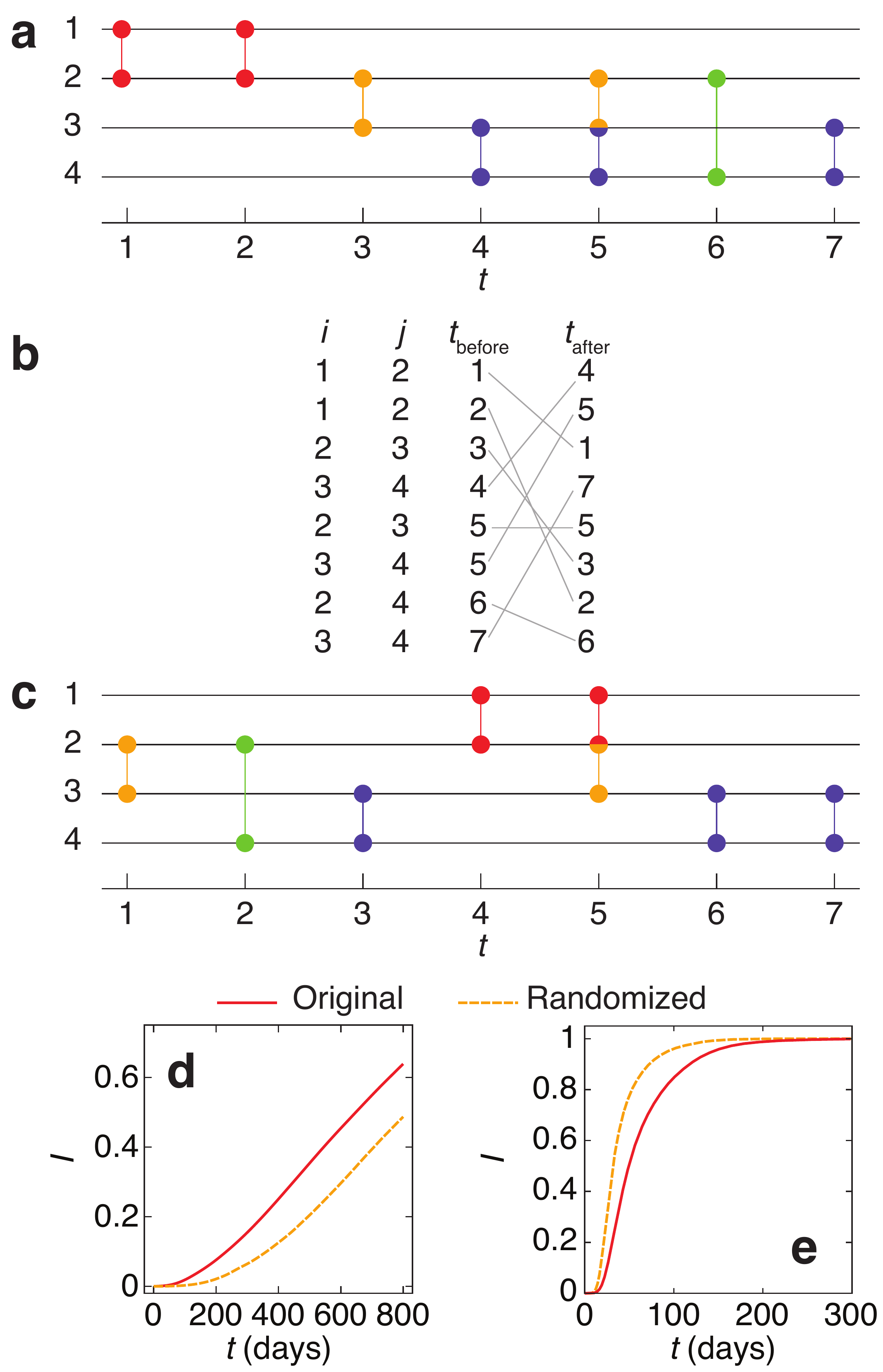}
\end{center}
\caption{Illustrating the shuffled-time-stamps scheme to randomize temporal networks. Panel (a) shows the original network from Fig.~\ref{fig:lossless}. Panel (b) shows how the randomization scheme operates on a contact-list representation of the data. In panel (c) we see the result as a nodal time-line plot. In panel (d), we see the effect of this type of randomization for susceptible-infectious (SI) spreading with 100\% infection rate. The plot shows the average number of infectious nodes $I$ as a function of time since the beginning of the data set $t$. The data comes from Ref.~\cite{rocha_comp_biol} and one can conclude that the order of event (that is destroyed by the randomization) speeds up the spreading. In panel (e), we see a corresponding plot for mobile phone data from Ref.~\cite{karsai_slow_small} where the conclusion is the opposite---spreading is slowed down, in the data, by the structure of the order of the contacts.}
\label{fig:randomization}
\end{figure}

\subsubsection{Shuffled time stamps}
 
To understand the effects of the order of events, one can randomly permute the times of contacts while keeping the network structure and the numbers of contacts between all node-pairs fixed~\cite{holme_reachability,karsai_slow_small,rocha_comp_biol,pan_saramaki,cardillo_evo,kivela_multiscale,miritello_dyn_strength,backlund_threshold,redmond_harrigan,starnini2012}. Algorithmically, this is very simple. Starting from a three-column list of contacts, one only needs to
\begin{enumerate}
\item Iterate through the contacts. Let $i$ be the current contact.
\item Take a random row $j$.
\item Swap the time stamps of rows $i$ and $j$.
\end{enumerate}
If identical contacts are forbidden, one would need to check that before swapping the time stamps (and iterate from step 2  until one finds $j$ to swap with without introducing such a consistence). Usually, such cases are so rare that one could ignore them and run the dynamic simulation one is interested in on a temporal network with duplicated contacts. This randomization scheme retains the overall number of contacts in the network during any period of time (i.e.\ preserving the overall daily and weekly cycles of activity in social data). It does destroy temporal effects like burstiness (at least some of it) and causal, triggering effects like person A calls person B because person C called A.

\subsubsection{Random times}

If one wants to understand the effects of cyclic patterns (daily patterns in human and biological data, weekly patterns in human data, etc.)~\cite{jo2012circadian,lahiri_berger_wolf_2008} one can compare the results from the shuffled-time-stamps randomization with a ``random times'' scheme. Here one replaces the time stamps with a random time, typically from a uniform distribution of the same duration as the original data. The additional effect from this randomization compared with shuffled time stamps explains the contribution from periodic patterns~\cite{holme_reachability,karsai_slow_small,pan_saramaki,backlund_threshold}. There is, however, a caveat for data sets where the number of present nodes is changing. For example the prostitution data of Ref.~\cite{rocha_liljeros_holme}, analyzed in e.g.\ Refs.~\cite{rocha_comp_biol,valdano_poletto,holme_liljeros,holme_info,fantulin_patient0}. In this data, the number of users active at a time $t$ (in the sense that they have been involved in a contact before $t$ and will be again after $t$) is growing until late in the sampling time. Assume that all agents have the same contact rates. Then, if one straightforwardly applies the above-described randomization, the contact rates of the nodes present only a short time will much lower than those spread out over most of the sampling time. If this effect is not desired, one could perhaps assign random times only to intermediate contacts over an link (contacts between the first and last). Since this would not remove the periodic patterns from the first and last contacts, there could still be biases if there are many links with just a few contacts (which is often the case). One could perhaps remedy this by adding a random number to the time of the first and last contact. However, there is probably no obviously best (or most principled) way to draw this random number. Which serves well to illustrate the difficulties of randomization techniques---just like centrality measures and community structure~\cite{newman_book,barabasi_book} there are many good ways of doing it, but none that is obviously, under all circumstances, the correct one.

\subsubsection{Random link shuffling}

The above two randomization schemes destroy temporal structures. One could of course neutralize topological structures as well. To do this, one can use a link shuffling method defined as follows
\begin{enumerate}
\item Pick an link  $(i,j)$ (sequentially) from the list of links.
\item Pick another link $(i',j')$ randomly.
\item With equal probability replace $(i,j)$ and $(i',j')$ by $(i,j')$ and $(i',j)$, or by $(i,i')$ and $(j,j')$.
\item If the move in the previous step created a self-link or multiple link, then undo it and start over from step 2.
\item Go to step 1.
\end{enumerate}
There is no statistical grounded criteria, known to the author, saying how many times to iterate the above rewiring procedure. Milo et al.~\cite{uniform} recommend 100 rewirings per link. Finally, one would randomly redistribute the contact sequences of individual links in the original temporal network, to the randomized one.

This randomization procedure destroys all network topological structures except the degree sequence of the original graph (of accumulated contacts). It preserves the overall activity level (or rather, and stronger, the set of times of contacts), and all statistics related to interevent times, like burstiness. The correlations between the overall activity of a node and the links it participates in are, on the other hand destroyed. This list of quantities preserved or destroyed could be made longer---triggered event sequences of e.g.\ person A calling person B calling person C would also be destroyed, unless they concern only two nodes (cf.\ ``ping-pong patterns''~\cite{zhao_etal,eckmann,jari_moro}). This highlights another slight  drawback of randomization schemes---it can be hard to untangle all their effects. However, it becomes less of a problem when one use increasingly wide randomizations to remove structure step-by-step.

\subsubsection{Time reversal}

Mostly to show the versatility of null-models, we mention Bajardi et al.~\cite{bajardi2011} and Donker et al.~\cite{donker_reverse} who investigate spreading on a network where time runs backwards. The idea is that it could estimate the number and importance of ``casual sequences'' where one contact triggers another that triggers another, and so on, into an outbreak cascade. As common with inference of causality, this method would rest on more assumptions than one would ideally want, nevertheless it shows a qualitatively different null model from the above.
 
\subsubsection{Poor man's reference models}

A difference with the time reversal null model, compared to the previously mentioned ones, is that it does not define an ensemble of temporal networks, but only one new temporal network. Sometimes computational constraints could make averaging over an ensemble impossible. Ref.~\cite{holme_liljeros} seeks to neutralize the effects of the distribution of interevent times over links (keeping the first and last contact intact, as mentioned above), but cannot sample an ensemble of randomized network. Instead they spread all the contacts between the first and last equally in time. This creates one network which arguable lacks the structure (broadly distributed interevent times), even though it lacks the same methodological basis (as it violates the maximum entropy principle) as a randomized network model. Ref.~\cite{holme_liljeros}, furthermore, explore the effects of the distribution of the time to the first contact, and from the last contact to the end of the sampling, in a similar manner---by moving all contacts to the beginning and end, respectively. Similar to the interevent times, this transforms a temporal network to another temporal network (not an ensemble) that lacks a structure, but instead of being distributed maximally randomly, it is just fixed to zero. As a stand-alone argument it would just be a stopgap measure, but Ref.~\cite{holme_liljeros} combines it with other ways of testing the effects of interevent times, times from the beginning to the first contact, and from the last contact to the end.

\subsubsection{Randomization---conclusions and practical advice}

As mentioned, we think randomization techniques are  powerful in temporal network studies, primarily as a method to analyze data sets (with the  benefit over generative null models that one starts the exploration at the data set itself). One could also use randomization for more abstract questions. Indeed, mentioned references like Ref.~\cite{holme_reachability,holme_liljeros,bajardi2011,karsai_slow_small,backlund_threshold} all uses randomization as intermediate steps towards a larger theory of how temporal network structure affects dynamic systems on the network.

The best methodology for randomization  is be to build a sequence of gradually more random ensembles. This could be done in several ways---see Refs.~\cite{holme_reachability,karsai_slow_small,backlund_threshold} for three different examples. Which way to choose is both question and system dependent, but it is important to make the randomization easy to interpret in terms of the structure it is supposed to destroy. This is, as mentioned, hard, and reading the literature one cannot help but thinking it would be helpful if someone worked out a systematic, principled way of doing this.

\subsection{Boundary conditions and extrapolations} \label{sec:boundary}

The range of possible structures and correlations in temporal networks is much larger than in static networks. That is a good argument for taking an empirical data set as your staring point and using randomization (Sec.~\ref{sec:random}) techniques, rather than tuning the structure of a generative model. The challenge with this approach is how to generalize the results to larger networks and longer times. Of course any kind of answer to this question involves some form of modeling of the data. One approach to compensate for a short sampling time $T$ is to use periodic boundary conditions~\cite{pan_saramaki,rocha_rw_cent,backlund_threshold,colizza_analytical,ogura_preciado}, i.e.\ by the time you reach the end of the sampling time $t=T$, you just start reading from the beginning $t=0$ but add $mT$ to the time for sweep $m$ of your temporal network. This method assumes that the ongoing link picture holds for the data (cf.\ Sec.~\ref{sec:ongoing}). The obvious problem with periodic boundary conditions is that they ignore temporal effects longer than the sampling time window. Worse yet, they can introduce new structures. If, for example, a node has a life time in the data of duration $\Delta t \ll T$, then periodic boundary conditions would introduce spurious interevent times of durations of about $T$. If the data, however, fits well to the ongoing link picture, then periodic boundary conditions are a good idea.  There are periodic temporal networks too, where periodic boundaries are exact~\cite{fujiwara,flocchini}---public transport systems, communication with satellites in low-Earth orbit, and security guard tours are three examples (from Ref.~\cite{flocchini}).

How to extend results for empirical data to larger populations is an interesting future challenge. Rocha et al.~\cite{rocha_comp_biol} add their temporal network data of sexual contacts in prostitution to a background, non-prostitution sexual network estimated from surveys. This approach, however, cannot be better than the coarsely modeled background data. Another approach that we have not seen evaluated, or used, yet is to sample random subsets of the nodes and by tuning their sizes and run the dynamic system on these. Then plot the quantities of interest as function of the size of the subsampled data (so the largest data point would be the entire data set itself). By this method, one could extrapolate results to larger sizes than the original network. However, this finite-size scaling method (straightforwardly applied) would introduce increasingly large biases the smaller the subnetwork is (simply because sampling half as many people is not the same as having a half as big population). Further research in this type of methods would be very interesting. Could one, for example, use empirical data as building blocks for arbitrary large, semi-empirical data sets without introducing fatal biases?

\subsection{Temporal link prediction}

Given a static network, assuming there are some missing links in it, \textit{link prediction} is the problem to rank pairs of nodes~\cite{lu_zhou} in order of their likelihood of being a missing links. In temporal networks, this is usually rephrased as to predict all links in the next time step, or further into the future. See Ref.~\cite{tempolinkpred} for a survey of this fairly large theme of mostly computer science. As an example of a method, Dunlavy et al.~\cite{dunlavy_temp_linkpred} (and several others) base their prediction methods on tensor factorization of the adjacency tensor. It would be interesting to predict missing contacts in a contact sequence. Given that an adjacency matrix does not have to represent an evolving network of accumulated contacts, this is a more principled generalization of link prediction to temporal networks (although it  makes less sense for computer science applications).

The slightly different problem of G\'enois et al.~\cite{genois} make a nice segue from this section to the next. G\'enois et al.\ study how to compensate for missing temporal links in empirical data for studies of spreading processes. Their method is based on constructing a weight matrix (where weight represents the fraction of all contacts that happened between two nodes). This means that they assume the ongoing link picture (cf.\ Sec.~\ref{sec:ongoing}) which could be a severe limitation~\cite{holme_liljeros,miritello_limited}. From the weight matrix, G\'enois et al.\ generate synthetic contacts that could be added to the original data to densify it.

\subsection{Generative models}

Generative models of temporal networks have a slightly different role than  in static networks. For classic network theory~\cite{barabasi_book} an important goal, especially in the early aughts, was to construct network models that generated some emergent structure (most commonly power-law degree distributions~\cite{ba_model}, but also e.g.\ community struc\-tu\-re~\cite{seceder}). For temporal networks there is no known, ubiquitous (or at least very common) structure that combines time and topology in a non-trivial way. On the other hand, there is a multitude of structures that can interact with, and affect, dynamic systems. The function of the models in temporal networks is thus more as tools to investigate the relation between structure and dynamics than to discover microscopic mechanisms. Although, as we will see, there are models of the latter kind too.

\begin{figure}
\begin{center}
\includegraphics[width=0.8\linewidth]{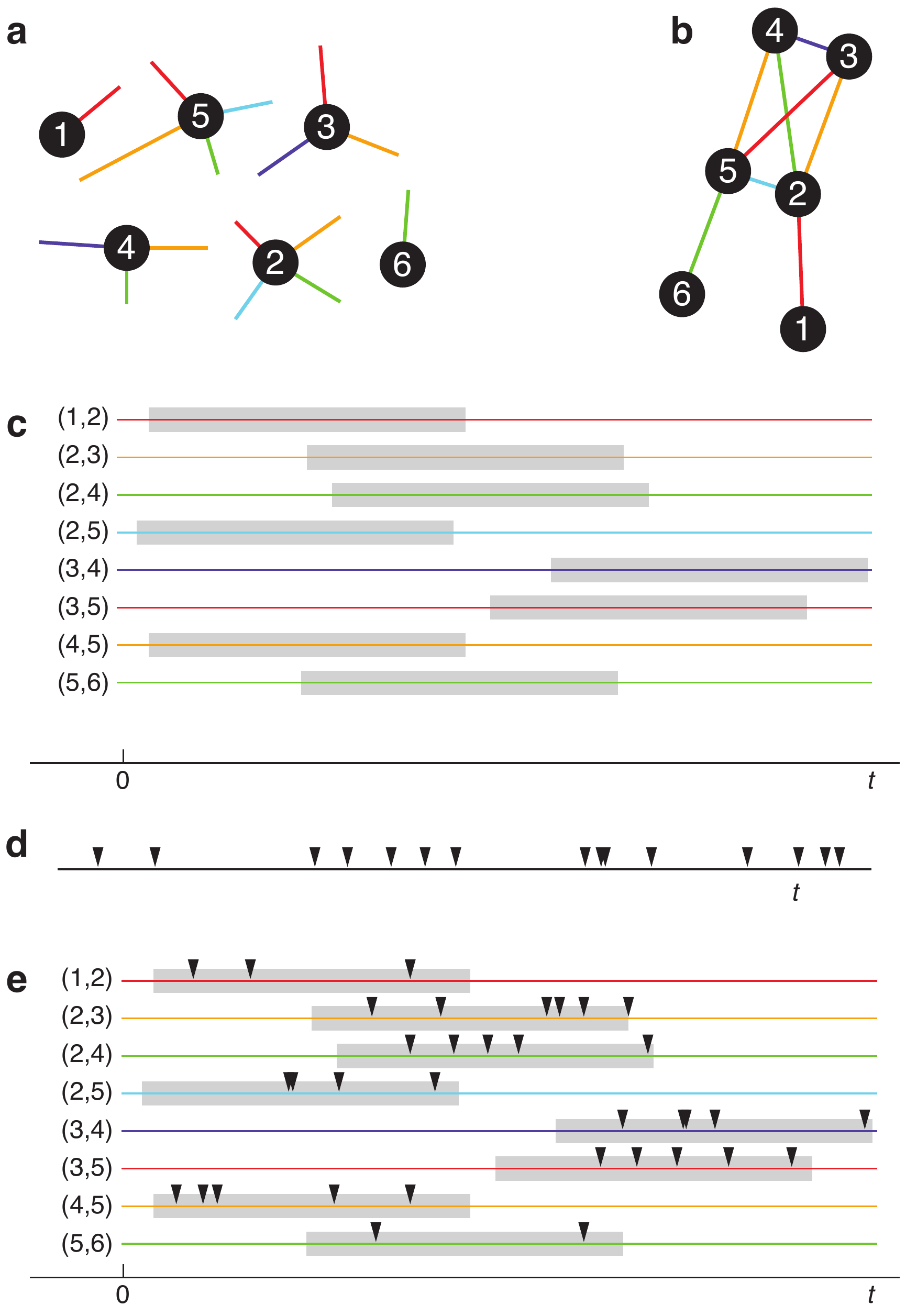}
\end{center}
\caption{Illustrating a simple generative model for temporal networks, used in Ref.~\cite{holme_comp_biol} and (almost) in Ref.~\cite{rocha_bursts}. First one generates a static network (technically a multigraph) from the configuration model by (a) drawing degrees from a probability distribution and (b) matching them up in random pairs. Then one generates active intervals for the links (also randomly, in this case all links being active the same duration), (c). Finally one generates a time series of  interevent times from a probability distribution (d) and rescales it and matches it to the active intervals.
}
\label{fig:model}
\end{figure}

\subsubsection{Static networks with link dynamics}

The most straightforward approach to generate a temporal network is to generate a static network from some model, and for each link generate a sequence of contacts. In the simplest approach, one would not let the contact generation be influenced by the network position of the links. For example, Holme~\cite{holme_comp_biol} uses the following protocol.
\begin{enumerate}
\item Construct a simple graph by first generating a multigraph by the configuration model~\cite{newman_book}, then removing duplicated links and self-links.
\item For every link, generate an active interval (when contacts can happen) from some distribution. Ref.~\cite{holme_comp_biol} uses a truncated power-law for the duration of the active interval, and a uniformly random starting time within a sampling time frame.
\item Generate a  sequence of contact times following some (bursty or not) interevent time distribution.
\item Wrap the contact time sequence onto the active intervals of the links. In other words, first rescale the sequence to the total time of the active intervals, then cut it in the same durations as the active intervals, and assign it to them.
\end{enumerate}
Adding correlations between the time and topology to this approach is fairly straightforward. This model is illustrated in Fig.~\ref{fig:model}.

Rocha and Blondel~\cite{rocha_bursts} use a method that is very similar to the above framework. The only difference is that  the active interval of a node  starts when that of another node ends. By this procedure, they generate networks where the number of active nodes is constant.

\subsubsection{Activity-driven networks}

Perra et al.~\cite{perra_activitydriven} propose a model of temporal networks that is simpler than the above. They use a graph sequence framework (Sec.~\ref{sec:graph_seq}), so $G_t$ denotes a simple graph at (the discrete) time $t$. The generation proceeds as follows:
\begin{enumerate}
\item Increase the time counter to $t$ and let $G_t$ be empty.
\item For every node $i$, make it active with a probability $a_i\Delta t$. Connect $i$ to $m$ other randomly chosen distinct nodes (active or not). Ref.~\cite{perra_activitydriven} uses a truncated power-law distribution for $a_i$.
\end{enumerate}
This model has been the ground for most analytical studies of processes on temporal networks~\cite{perra_rw,karsai_perra,liu_metapop,liu_controlling,starnini_pastor,issue:perra,han_sun} and emergent static network structures~\cite{starnini_topo}. Jo et al.~\cite{jo_perotti} study disease spreading on a similar (but not identical) temporal network model. Laurent et  al.~\cite{issue:marton} extend the original activity-driven model to include memory effects and triadic closure~\cite{thebook:higham1,li_clustering}. Yet an extension of the activity-driven model was proposed by Moinet et al.~\cite{moinet}. Their model seeks to include aging effects, in particular for scientific collaboration networks (cf.\ the long-term structures discussed in Sec.~\ref{sec:linkturnover}). In a final extension Sunny et al.~\cite{sunny} incorporate life times of links (including non-Markovian models that need to be solved numerically).

\subsubsection{Face-to-face interactions and communicators}

Starnini et al.~\cite{pastor_face2face} developed a model of temporal face-to-face networks. This is naturally a spatiotemporal network. Technically, their model is a two-dimensional random walk model where there the chance of walking closer to a node $i$ is proportional to an increasing attractiveness $a_i$.  The more attracted a walker is to its neighbors, the slower its walk becomes. Finally, they also model the agents as having active and inactive periods that they transfer between with the same  probability every time step. The authors motivate the last step by observations (that the people observed in e.g.\ Ref.~\cite{sociopatterns_hospital1} do not always socialize, even though they could). Zhang et al.~\cite{li_b} propose a slightly more elaborate model, but with an abstract representation of space, for the same problem.

Mantzaris and Higham~\cite{higham_dynamic_communicators} propose a model for communication in an online setting motivated by the observation that some individuals are much more central in a temporal sense than they are in an aggregated static network. Their method is somewhat similar to the above in that it assigns an intrinsic trait value to the nodes. Then it proceeds by assigning random communication partners to a node by a basal rate and a positive feedback mechanism. A more statistics oriented model of communication in social networks  can be found in Raghavan et al.~\cite{raghavan}, while Hsu et al.~\cite{hsu_clock} take a more mechanistic approach towards a similar goal.

\subsubsection{Link-node memory models}

Vestergaard et al.~\cite{vestergaard_how} propose a model where both nodes and links are activated by temporal (non-Markovian) effects. In their model, links can be active or inactive (like the above mentioned activity-driven model). Apart from their state, a link is characterized by the time $\tau_{(i,j)}$ since the last time it changed state. Similarly, a node $i$ is also acts depending on the time $\tau_i$ since it last was involved in a contact. The network is initialized to $N$ nodes and all links  inactive. An active link is inactivated with a rate $z f_\mathrm{link}(\tau_{(i,j)})$ ($z$ is a control parameter). A node can initiate (activate) a link with probability $b f_\mathrm{node}(\tau_{i})$. This new link is chosen from the nodes $i$ is currently not in contact with with a probability $\Pi_\mathrm{node}(\tau_j)\Pi_\mathrm{link}(\tau_{(i,j)})$. Where the memory effects enters the model through the ``memory kernels'' $f$ and $\Pi$. Vestergaard et al.\ give these a power-law decay form and shows that the with a proper choice of exponents, their model can recreate the statistics of many empirical temporal networks.

\subsubsection{Self-exciting point processes}

Masuda et al.~\cite{thebook:masuda} and Cho et al.~\cite{cho} use a Hawkes process with the same objective as Starnini et al.~\cite{pastor_face2face} above. Masuda et al.\ observe that there is a positive correlation between consecutive interevent times in empirical data that cannot be modeled by  interevent times alone. Their model works by defining an event rate at time $t$ as
\begin{equation} \label{eq:hawkes}
v+\sum_{i:t_i\leq t}\phi(t-t_i)
\end{equation}
where $\phi$ is an exponentially decreasing memory kernel (zero for negative arguments to respect causality) and $v$ is a basal event rate. Even with an exponentially decreasing kernel, the interevent time distribution becomes heavy-tailed. The model also creates positive interevent time distributions, but not the gatherings that can appear in real face-to-face networks (Sec.~\ref{sec:proxi}). Cho et al.\ extend this framework by including spatial effects. Zipkin et al.~\cite{zipkin} make a comprehensive study of point-process models of social-network interaction but without a coupling to the topology.

Like the above, Colman and Vukadinovi\'c Greetham~\cite{colman_greetham} propose a model of temporal networks founded on the theory of stochastic point processes. In their setup, a node forms and break links based on a Bernoulli process with memory. Similar to the Hawkes process mentioned above, the probability of an event between $i$ and $j$  increases with the number of recent events that happened between $i$ and $j$. More precisely, Colman and Vukadinovi\'c Greetham take the probability of a link to activate or deactivate at time $t$  to be proportional to the number of such events in a time window of a certain duration ahead of $t$. For the model we just sketched, the authors derive an emergent power-law interevent time distribution.

\begin{figure}
\begin{center}
\includegraphics[width=\linewidth]{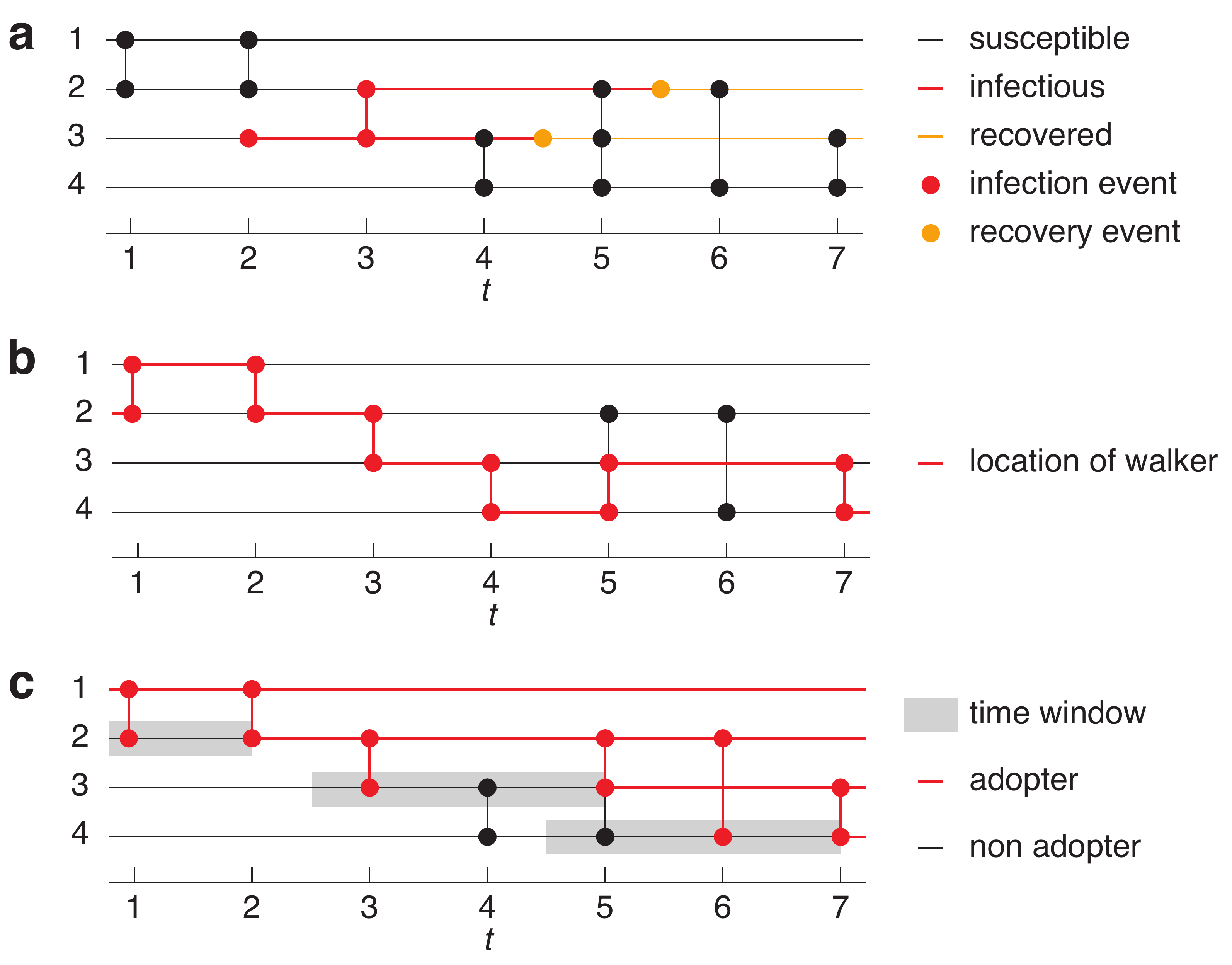}
\end{center}
\caption{Illustration of three dynamical systems on temporal networks. Panel (a) shows a susceptible-infectious-recovered model with a disease duration of 2.5 time steps. The outbreak starts at node 3 at time 2 and reaches one other node. There is a potentially contagious event between nodes 3 and 4 at time 4, but in this example chance made it not contagious. Panel (b) shows a greedy walk starting at node 2 at the beginning of the sampling period. A greedy walk follows every contact away from the node where the walker is. Panel (c) illustrates one of the temporal threshold models studied in Ref.~\cite{karimi_holme}. A node becomes adopter if it is exposed to (i.e.\ in contact with an adopter) more than twice within a backward time window of 2.5 time units.}
\label{fig:processes}
\end{figure}

\section{Dynamic systems on temporal networks}

The study of any kind of network is usually done with some dynamic system in mind. The network is just the infrastructure for the dynamics. As already mentioned, a central theme in temporal networks is to relate the structures in both time and topology, to dynamic systems. The types of dynamic systems of interest can be different compared to static networks~\cite{porter_gleeson}---those that are naturally much faster than the underlying network, like traffic on a road network, become less interesting; those that are  dependent on the timing of the contact could be meaningless in static networks.

\subsection{Walks}

A walk is a process where one (or possibly more) units (\textit{walkers}) move with the contacts across a temporal network. In other words, the  walker is always located at one and only one node. If it is at $i$ at time $t$, and there is a contact $(i,j,t)$ then the walker can go to node $j$. If the walker makes the move or not could be decided in many different ways. Usually the decision process has a random component, making it a \textit{random walk}~\cite{lambiotte_burstiness,speidel_rw,baronchelli_resolution,delvenne_timescales,luis,rocha_rw_cent,sousa_pastor}.  The limit of certainty defines \textit{greedy walks}~\cite{issue:saramaki_holme}. 

Random walks are a very fundamental stochastic process with applications in many areas, from physics to economy, biology and social science~\cite{havlin_rw}. For temporal networks, it is the dynamic system that is easiest and most straightforward to adapt from static networks. One do not need to make any of the subtle technical decisions mentioned for epidemic models below. On the other hand there are perhaps not so many real phenomena modeled by random walks on temporal networks. Our only motivating example is Ramiro et al.~\cite{ramiro_rw} who argue they could be a lightweight communication protocol, and perhaps the paths of travelers in air-travel data~\cite{rosvall_memory}. The reason walks do not model many real systems is that things spreading on temporal networks are usually not conserved, or it is (like for distributed computing) the thing spreading that defines the temporal network. For this reason, the classical quantities of random walk theory (e.g.\ first passage time) are also not as central as for random walks on lattices or static graphs---but they do exist~\cite{starnini2012,perra_rw,issue:lambiotte,speidel_rw}. Instead people either use walks as probes of the system's burstiness~\cite{starnini2012,luis}, time scales~\cite{delvenne_timescales,baronchelli_resolution} or other temporal patterns~\cite{issue:saramaki_holme}. Another use of random walks (alluded to above) is to use them as a basis for a generative model of temporal networks~\cite{barrat_itineraries,speidel_rw,hoffmann_rw,thebook:lambiotte,masuda_klemm}. Delvenne et al.~\cite{luis} separate temporal and topological effects in random walks on networks. Where one typically use a fully connected underlying topology (except Barrat et al.~\cite{barrat_itineraries} who start from a weighted static network). Memory network~\cite{rosvall_memory,scholtes_causality} are also modeling temporal networks as random walks (beyond a first order Markov process).

\subsection{Epidemic models} \label{sec:epi}

The spread of infectious disease is one the most natural types of dynamic processes on temporal networks~\cite{masuda_holme_rev}. The framework for modeling epidemic outbreaks is well established since decades. In \textit{compartmental models}, one divides the population into states (classes, or compartments) with respect to the disease, and assign transition rules between these classes. The four most common states are \textit{susceptible} (individuals that do not have the disease, but can get it), \textit{infectious} (who have the disease and can spread it), \textit{recovered} (who cannot get or spread the disease), and \textit{exposed} (who are infected but cannot yet infect others). The infection event is always (as far as we know) between a susceptible and an infectious and is usually the only transition that involves a human-to-human interaction. The second, equally important, ingredient in epidemic modeling is a model of the contact patterns, i.e.\ a model or data set describing who meets whom, when and (sometimes) where. Temporal networks fit well for this job, and no wonder much of the temporal-network theory have been developed with infectious diseases in mind. Hethcote's Ref.~\cite{hethcote} is our favorite introduction to the classic theory of epidemiology; Keeling and Eames~\cite{keeling_eames} is an excellent introduction to static networks as models of contact structure; and Ref.~\cite{masuda_holme_rev} gives a brief review of temporal network epidemiology.

Simulating disease spreading on temporal networks is easy. Or rather, the straightforward way of doing it is also quite efficient. The reason is that we assume the entire contact sequence is an input. There is thus no reason to model the contacts, like disease simulation algorithms on static graphs need to do~\cite{holme_logistical}. Still, there are many technical issues to consider when modeling disease spreading on temporal networks. First, how one should model the exposed and infectious stages. In traditional mathematical epidemiology, one assumes a finite chance of exiting these states every unit of time. This means that the duration of those states will be exponentially distributed, which is not in agreement with data~\cite{lloyd}. The advantage is that (since one time step is independent from the next) the analytics gets much simpler. The other option is to assume these stages have constant durations, which is also a big simplification. The advantage with this approach, except that it is more realistic, is that it the programs become a bit faster and sometimes more compact. Next, practical consideration is what to do if there are several contacts involving a node during the same time step. The solution depends on the data set and how much you would like to stick to the assumption that the contagion is instantaneous. One principled way is to sample the contacts during a time step in a random order, another is to not allow a contact to spread the disease the same time step they got it (effectively inserting an extra exposed state). The last, and perhaps biggest, technical issue is the boundary condition. Basically, you would like to assume that spreading can only happen within the data, while at the same time, scale up the results to a much larger population than your data set. G\'enois et al.~\cite{genois} make a pioneering attempt in resolving this issue by proposing a resampling method to extrapolate the results to large populations. There are  many technical challenges left to solve, and many redundancies to exploit to tune the algorithms. Vestergaard and G\'enois~\cite{temporal_gillespie} proposes an extension of the Gillespie algorithm (that, rather than simulating an infection event by random numbers, calculates the time to the next successful infection). It is worth noting that, the most canonical compartmental models---the susceptible-infectious-susceptible (SIS) and susceptible-infectious-recovered models (SIR)---are two-pa\-ra\-me\-ter models in temporal networks, but for both static networks and fully connected models, they are effectively one-parameter models. This changes the analysis more than one could first expect, and it is easy to miss some effect by analyzing them as if they were one-parameter models.

% how structure affects
Many papers in temporal network epidemiology (especially in the early days) showed the big differences that adding the time component to the contact structure can make~\cite{riolo_etal,moody,fffng,bramson,liu_metapop,rocha_comp_biol,holme_info,zhu,horvath,machens,ren_epi,rocha_epi}.
One line of research in temporal network epidemiology concerns how structural aspects affect disease spreading. The progenitors of this theme was the early HIV researchers who compared serial monogamy to concurrent partnerships~\cite{morris_kretzschmar_concurrent}. In the recent wave of temporal network research an early finding was that burstiness---a broad distribution of time between contacts---slows down disease spreading~\cite{karsai_slow_small,min_goh_vazquez,horvath}. Ref.~\cite{rocha_comp_biol}, however, found that time  shuffling (destroying burstiness) is slows down SI spreading.  The reason for this observation, argues Ref~\cite{holme_liljeros}, is that the number of active nodes in the data set of Ref.~\cite{rocha_comp_biol} keeps growing throughout the sampling time. This is a condition for the ongoing-link picture to hold, which is assumed by papers arguing that bursty link dynamics slows down spreading (see Sec.~\ref{sec:ongoing}). Rocha and Blondel~\cite{rocha_bursts} continue this analysis with different spreading, scenarios and structures on model temporal-networks. Briefly speaking, they conclude that it is hard to generalize the relationship between structure and spreading statistics over different compartmental models. A quite similar result is presented by Sun et al.~\cite{issue:perra} who show that memory effects in model temporal networks can increase the epidemic threshold of the SIR model but lower the threshold in the SIS model. Karsai et al.~\cite{karsai_perra} show that, quite counterintuitively, strong links (paris of nodes with many contacts) can impede a SIR spreading process. Sun et al.~\cite{issue:perra} make this result more nuanced by showing that in SIS spreading the strong links can prolong the outbreak. Analytical studies are rarer than computational (no wonder---adding a level of complexity to network mathematics makes it really challenging), but Zhang and Li~\cite{li_sis} and Valdano et al.~\cite{colizza_analytical} derive the epidemic threshold for a SIS model using a rate equation and spectral method respectively. There are also a handful studies using Perra et al.'s~\cite{perra_activitydriven} activity-driven and similar models to study epidemics~\cite{karsai_perra,liu_metapop,sunny,liu_controlling,issue:perra,han_sun,jo_perotti}.

%connecting to tradition
There is a number of papers in temporal network epidemiology that tries to connect to the traditional theoretical epidemiology. One example is studies analyzing the \textit{basic reproductive number} $R_0$ in temporal networks. $R_0$ is defined as the expected number of people a first infectious person would infect in a population of only susceptibles. This is an estimator of the spreading speed of a specific disease in a specific population, that many (both medical and theoretical) epidemiologists are so used to that they want to see it estimated, even though  direct quantities---prevalence (number of infectious individuals) and incidence (number of infection events)---are not harder to estimate than $R_0$. That much said, $R_0$ is an important bridge between temporal-network epidemiology and classical theoretical epidemiology (such as discussed in Ref.~\cite{hethcote}). Rocha and Blondel~\cite{rocha_bursts} estimate $R_0$ for many spreading scenarios in model temporal networks. Holme and Masuda~\cite{holme_masuda} discuss how $R_0$ can fail as an estimator for the final outbreak size. Consider two diseases with per-contact transmission probabilities $\lambda_1$ and $\lambda_2$ and disease durations $\delta_1$ and $\delta_2$. In principle, the outbreak size grows with both $\lambda$ and $\delta$, so there are naturally parameter values $\lambda_1>\lambda_2$ and $\delta_1 <\delta_2$ that gives the same $R_0$, but these do not have to give the same outbreak size. In general, there are parameter values where disease 1 has a larger $R_0$ but disease 2 has a larger estimated outbreak size. Holme and Masuda~\cite{holme_masuda} discuss the conditions for this, somewhat paradoxical, situation to happen and find that sometimes this can be explained by topological properties, sometimes because of temporal structures. Another way to connect the bridge between temporal networks and static networks is to ask how to best reduce a temporal network to a static network (i.e.\ so that the static network encodes as important epidemiological information as possible). The simplest way is to assume a time-windowing procedure (see Sect.~\ref{sec:window}) and ask how to choose the optimal time-window~\cite{bernhardsson,liljeros_giesecke_holme,holme_comp_biol}. The rough answer (as far as epidemic spreading goes) is that one should set the beginning of the time window as close as possible to the beginning of the epidemics and the end of it to match the time scale of the spreading~\cite{holme_comp_biol}. Ref.~\cite{holme_info} investigates how the additional,  information of temporal networks affect the predictability of outbreak sizes as a function of when the outbreak is observed.

%preventing
Another line of research concerns the identification of influential spreaders. One version of this is the \textit{vaccination problem}---imagine you can immunize (or in other ways lower the impact with respect to spreading) a fraction $f$ of a population, then how would you chose them. One needs to make further assumptions about what kind of information that is accessible, and how reliable it is, which make this a rich question. Lee at al.~\cite{sm_vacc} assume individuals can name who they have been in contact with (such that the disease could spread) and when the contacts happened. They suggest to sample people at random and then vaccinate their most recent contact, which both increases the chance of vaccinating active people  in general and those in a current burst of activity. Starnini et al.~\cite{starnini_vacc} extend this research, assuming global knowledge about the contact could be obtained and try protocols like  vaccinating nodes with the highest degree, or betweenness, over a time window before the vaccination. G\'enois et al.~\cite{genois_vacc} also seek vaccinees by a form of estimated betweenness. Specialized to corporate employees, they propose to first vaccinate people who share their time between different departments of the company. Osawa and Murata~\cite{osawa_murata} identify vaccinees by growing a cluster in a greedy algorithm. Habiba et al.~\cite{berger_wolf_vacci} assume full knowledge of the contacts and investigates vaccination by various temporal-network centrality measures. Salath\'e et al.~\cite{salathe} also assume global knowledge and argue that temporal patterns are important for preventing the spread of influenza (with proximity data from an American high school). Toth et al.~\cite{toth} present a similar study of influenza spreading in a real network of elementary and middle school children. Ultimately, the vaccination problem is a prediction problem. One needs to decide whom to vaccinate to prevent future disease spreading based on the knowledge at the time of the vaccination (which can only be about the past). Valdano et al.~\cite{valdano_poletto} discuss this issue in more general terms and connects it to the presence of persistent structures (Sec.~\ref{sec:persistent}). Mantzaris and Higham~\cite{thebook:higham2} argue that communicability is a good predictor for how easily an outbreak can spread on a temporal network. Gauvin et al.~\cite{latent_gauvin} use their tensor factorization method from Ref.~\cite{gauvin_panisson} to identify temporal subnetworks that could contain a disease within itself longer than expected by chance (thus acting reservoir for the disease).

Even though most applications of temporal networks to epidemic models concerns human diseases, there is a particularly active subfield interested in disease spreading in livestock~\cite{vernon,gates_woolhouse,scharrer,valdano_poletto,konschake_components}. The reason is twofold. First, there are good data sets on animal transportation. Second, a contact is fairly well-defined for such temporal networks. (Which is also a reason sexual networks~\cite{watts_may,morris_kretzschmar_concurrent,rocha_bursts,kretzschmar_etal,rocha_comp_biol} and hospital-transmitted infections are frequently studied in the temporal-networks literature~\cite{cori,liljeros_giesecke_holme}.)

\subsection{Opinion and information spreading} \label{sec:infospread}

Except disease spreading, the other major class of one-to-many spreading phenomena in temporal networks of human contacts is that of rumors, opinions, information, etc. By now it is a quite a clich\'e that these spreading phenomena work differently than infectious diseases (``viral videos don't spread like viruses''). We have learned from studies of spreading on social networking services that individual behavior is very diverse. Not only do people have different levels of activity, they could also follow completely different mechanisms~\cite{romero,DeMartino2015}. The term ``complex contagion'' refers to processes where a compartmental model of disease spreading is not enough, which is often, but not always, the case for information spreading. There is some ambiguity here---Ref.~\cite{guille} defines complex contagion as when the content of what is spreading affects the spreading (like that one would have to be exposed more to an idea the more controversial or contentious it is to adopt it).

Maybe the simplest type of complex contagion models are threshold models. These assume that an individual adopts an idea when the exposure is over a threshold. For temporal networks, this adds complexity compared to disease spreading models because, clearly, older influence matters less than newer, so one has to decide how to down-weigh the older contacts. Refs.~\cite{karimi_holme,thebook:karimi} adapts Watts's cascade model~\cite{watts_thres}  to temporal networks by counting only contacts within a moving window. If a large enough fraction, or number, of the contacts within this window is to adopters (i.e.\ individuals carrying whatever is spreading on the network), then an agent would adopt it. Ref.~\cite{taro_holme_threshold} uses a slightly different threshold model, where the importance of a contact with an adopter decays exponentially in time. All these three studies~\cite{karimi_holme,thebook:karimi,taro_holme_threshold} conclude that there are situations where burstiness would have slowed down spreading in epidemic-type models, when it would be accelerated for threshold models. The intuition is that the contagion might need a burst to overcome the threshold. Ref.~\cite{backlund_threshold} also uses a moving time window (like Ref.~\cite{karimi_holme}), but calculates the ratio of adopters among the number of neighbors in the network of aggregated contacts. Thereby they weigh exposure coming from different neighbors heavier than repeated information from one neighbor. De Martino and Spina~\cite{DeMartino2015} extend the SI model to include individual contact rates and can thereby match empirical spreading data more accurately. Finally, we mention Ref.~\cite{michalski} that investigates methods to seed a temporal network (with applications to word-of-mouth marketing) for optimal spreading by a threshold type dynamics.

An even simpler type of opinion spreading model than threshold models is the voter model. In this model random nodes copy the opinion of random neighbors. Although very popular in static networks, we only know two works~\cite{hoppe,thebook:eguiluz} studying it on temporal networks. In this study, the temporal network is a result from a mutual selection process. Comparing the voter model to other dynamic models on empirical data sets would be interesting. Nishi and Masuda~\cite{nishi_masuda} study another model of opinion spreading based on social balance theory. This theory concerns networks with both positive and negative links. A positive link means that two actors like or benefit from each other. The theory states that triangles with an odd number of negative links are not stable. This principle could straightforwardly be changed into a network evolution model. Nishi and Masuda take another approach and assume the interaction to be given, but the link signs to evolve toward a more socially balanced state. They argue that temporal fluctuations slow down the time to global social balance.

\subsection{Percolation, error tolerance and attack vulnerability}

Compared to random walks, and spreading processes, there has been relatively little studies of percolation theory on temporal networks. Otherwise, percolation is well studied, especially on  regular lattices~\cite{durrett} but also on static networks. Simply speaking, percolation theory colors the nodes of a network black and white by some random process, then asks the probability that there is a connected path of black nodes reaching from one side of the network to the other. Originating in geophysics, percolation theory has also been used in material physics, and many types of interdisciplinary applications. Starnini and Pas\-tor-Sa\-tor\-ras~\cite{starnini_pastor} is the only paper we are aware of that is fully devoted to percolation on temporal networks. Their main point is to map an SIS process to a percolation problem, which they test on the generative temporal-network model of Ref.~\cite{perra_activitydriven}. P\'osfai and H\"ovel~\cite{posfai_structural} study percolation as a step towards studying network controllability (see also Ref.~\cite{li_controllability}). Miritello,  Moro and Lara~\cite{miritello_dyn_strength} map a spreading problem to percolation. Indeed, the SI model~\cite{rocha_comp_biol,bayhan_hoplimit,rocha_bursts,barrat_activity_clocks,albano_ex_intrinsic} is equivalent to invasion percolation.

One way of interpreting percolation theory in networks is as a way to estimate the functionality of a networked system in the presence of failed components. If there is a connected cluster spanning the network despite a random fraction of the nodes broken, then the network, as a whole, is still considered working. Alternatively, one can think of an adversary that deliberately wants to break the networked system. In static networks, heterogeneities are known to improve the error tolerance but worsen the attack vulnerability~\cite{ba_attack}. Trajanovski et al.~\cite{tarjanovski} and Sur et al.~\cite{sur_niloy} arrive at a similar conclusion for temporal networks. Trajanovski et al.\ test a variety of attack scenarios. Sur et al.\ specialize in networks with a pronounced temporal community structure. Note that the attack vulnerability problem is (if not equivalent, at least) very similar to identifying influential spreaders in information spreading (Sec.~\ref{sec:infospread}) or nodes to vaccinate to stop disease spreading (Sec.~\ref{sec:epi}).

\subsection{Synchronization}

There are  conspicuously few papers that take a temporal network as input and asks how fast, or well, a network of oscillators can synchronize. A paper about synchronization like Ref.~\cite{perra_rw} is about random walks seems like a low-hanging fruit. The closest ones we are aware of are Buscarino et al.~\cite{buscarino} who investigate synchronization on a network that evolves in an uncorrelated random way; Lee et al.~\cite{lee_lee} who study a periodically changing underlying network; and Kohar et al.~\cite{kohar_sync} who randomly rewire the network during the time evolution of an oscillator system. We briefly mention neuroscience where synchronization is an important concept. For brain networks (as mentioned in Sec.~\ref{sec:brain}) synchronization itself defines the contacts (Ref.~\cite{bassett_xlinked} and other papers) and is thus not a dynamic system on the network, but there could be more practical motivations for synchronization studies in this direction.

\subsection{Evolutionary games}

A very well studied type of dynamics on static networks is game theoretical models~\cite{szabo_fath}. These address the evolution of cooperation in a population of egoistic individuals, the evolution of populations that compete for a common resources, and many other scenarios where there is a conflict between short-term individual interests and longer-term interests of the entire population. Game theory models typically assign strategies (usually just labels) to the agents. Then, based on the interaction network and the strategies, they calculate a payoff for all the agents (that they seek to optimize). 

To our knowledge, there is only one paper studying game dynamics on temporal networks---Cardillo et al.~\cite{cardillo_evo}. They base the study on human proximity data that they aggregate over time slices. They run the game dynamics over a time window, calculate the payoff and update the strategies (by copying strategies from random neighbors, but with a higher chance of copying a strategy of a successful neighbor). Cardillo et al.\ focus most of their analysis on the effects of the time window size and the temporal network structure (analyzed by randomization null models).

\section{Discussion and future outlook}

We hope this expos\'e has given you an overview the state of temporal networks as of summer 2015. The field has evolved tremendously in the last five years. When we wrote Ref.~\cite{holme_saramaki_rev}, our feeling about the field was quite different. We thought the field was waiting for its scale-free degree distribution---some very common structure, involving both time and topology, that called for a non-trivial explanation. This, we thought, would lead to a flurry of activity in generative models. There have indeed been many generative models proposed, but not to explain a newly discovered structure. The focus have been more on what is spreading or diffusing on the networks, rather than the temporal networks themselves. This does not necessarily mean that there is no future for temporal-networks metrics. We cannot help but thinking researchers have, so far, been too locked into thinking like static network theory.  There could be other types of structures, very different from established ones in both static network theory and the theory of time series. In particular at the mesoscopic scale, we can imagine meaningful temporal network structures that are neither cohesive subgraphs, nor core-periphery structures, nor bursts, etc. There has also been a rather large amount of analytical works (even though the field is, on average, more computational than static network theory). This is of course a good thing, but we wish the analytical minded authors would stay data driven, and model  structures that: first, are observed in empirical data; second, can affect the dynamic system of interest.

Assuming the current trends will continue, we will see temporal network research  diverging from static network science and become even more temporal. There is plenty of room for such developments. For static networks, it was early established that degree distributions~\cite{barabasi_book} were the most fundamental statistics. Even for purely temporal quantities of a temporal network, there is no obvious such structure. In the wake of Ref.~\cite{barabasi_bursts}, there has been a lot of focus on interevent times. But quantities such as the time nodes or links first enter the data, the time between the first and last contact, the average time between their contacts, etc.\ are not more complicated and, at least under some conditions, more important for spreading processes~\cite{holme_liljeros,holme_masuda}.

We also anticipate new applied areas to discover temporal networks as a modeling framework. The last five years, neuroscience (more specifically, brain science) has started to embrace temporal network methods (Sec.~\ref{sec:brain}). There have been some temporal network papers in animal behavioral science  (Sec.~\ref{sec:aniprox}) and  ecology (Sec.~\ref{sec:econwk}), but we anticipate more studies in these fields. Especially since there are several long-term ecological studies that presumably have good temporal network data (see e.g.\ \url{http://www.lternet.edu/}).

A straightforward way of finding new research questions if of course to add yet a level of complexity (just like temporal network studies once spawned from network science). The most natural way would be to add either different types of links to get temporal multiplex, or multi-layer, networks~\cite{boccaletti_rev,kivela_rev}, or space to get  spatiotemporal networks. Williams and Musolesi~\cite{williams_musolesi} and Ref.~\cite{george_kim_spatiotemporal} discuss how to generalize many concepts (centrality etc.)\ to when space is added to temporal networks (or time to spatial networks). Sarzynska et al.~\cite{sarzynska} present a model to generate null models for spatio-temporal networks of interacting agents.

An area where we already in Ref.~\cite{holme_saramaki_rev} anticipated more activity  is visualization. As illustrated in Fig.~\ref{fig:graphical_mijin} and alluded to in Secs.~\ref{sec:static}, \ref{sec:timeline}  and \ref{sec:film}, temporal networks lack the intuitive visual component of static networks. Probably this is a fundamental property that cannot be completely altered, but  there should be better visualization methods than we have now. Highest on our wish list is a method that both simplifies some structures (cf.\ Refs.~\cite{seceder,tantipathananandh_etal,rosvall_alluvial}) and keeps (at least some) of the time-respecting paths (maybe at the cost of not having time on the abscissa). We should mention Zaidi et al.~\cite{zaidi_viz} who present some other ideas how to visualize temporal networks than we discussed here.

Finally, another direction, we would love to see more research in (mentioned in Sec.~\ref{sec:boundary}) is how to extrapolate results from e.g.\ spreading studies on empirical networks to larger populations. We believe there must be other methods (to be discovered) of resampling the original data, or scale up the results of such studies.

\subsection*{Acknowledgments}

PH was supported by Basic Science Research Program through the National Research Foundation of Korea (NRF) funded by the Ministry of Education (2013R1A1A2011947). We thank Des Higham, Xiang Li, Zohar Nussinov, Leto Peel, Mason Porter,  Jari Saram\"aki and Ingo Scholtes for helpful feedback.

\end{document}